\documentclass[a4paper]{article}
\usepackage{empheq,etoolbox}
\patchcmd{\subequations}
  {\theparentequation\alph{equation}}
  {\theparentequation.\alph{equation}}
  {}{}

\usepackage{stackrel}
\usepackage{subfigure}
\usepackage{amssymb,amsthm,amscd, amsbsy, array}
\usepackage{amsmath,bm,cite,color}
\usepackage{tikz}
\usepackage{graphicx}
\newcommand{\mathsym}[1]{{}}

\usepackage{epsfig}
\usepackage{epsf}
\usepackage{rotating}
\usepackage{color}
\usepackage{hyperref}
\newfont{\tenmsb}{msbm10 scaled\magstep1}

\let\ssection=\section\renewcommand{\section}
{\setcounter{equation}{0}\ssection}

\newcommand{\half}{{\scriptstyle{\frac{1}{2}}}}

\newcommand{\cP}{{\mathcal P}}
\newcommand{\cO}{{\mathcal O}}
\newcommand{\lf}{\left (}
\newcommand{\lfq}{\left [}

\newcommand{\rg}{\right )}
\newcommand{\rgq}{\right ]}

\def\smallover#1/#2{\hbox{$\textstyle{#1\over#2}$}}
\def\beq{\begin{equation}}
\def\eeq{\end{equation}}
\def\beq{\begin{equation}}
\def\eeq{\end{equation}}
\def\bea{\begin{eqnarray}}
\def\eea{\end{eqnarray}}
\newcommand{\av}[1]{\langle #1 \rangle}

\def\lf{\left(}
\def\rg{\right)}

\def\lgr{\left\{}
\def\rgr{\right\}}
\def\p{{\partial}}
\def\p{{\partial}}

\def\cT{{\mathcal T}}

\def\*{{\star}}

\newcommand{\FG}[1]{{\color{blue} #1}}

\newtheorem{theorem}{Theorem}

\newtheorem{proposition}[theorem]{Proposition}
\newtheorem{remark}[theorem]{Remark}

 \usepackage[final]{changes}

\definechangesauthor[color=green]{vito}
\definechangesauthor[color=purple]{3}
\definechangesauthor[color=blue]{2}
\definechangesauthor[color=red]{1}

\usepackage{soul}
\usepackage{enumerate}

\begin{document}

\title{Integrability properties and multi-kink solutions 
\\of a generalised  Fokker-Planck equation }

\author{
	Francesco Giglio$^{\; a}$\footnote{{\tt email: francesco.giglio@glasgow.ac.uk}}, 
	Giulio Landolfi$^{\; b,c}$\footnote{{\tt email: giulio.landolfi@le.infn.it, giulio.landolfi@unisalento.it}}, 
	Luigi Martina$^{\; b,c}$\footnote{{\tt email: luigi.martina@le.infn.it, luigi.martina@unisalento.it}},
    Andrea Zingarofalo$^{\; b}$\footnote{{\tt email: andrea.zingarofalo@studenti.unisalento.it}}
	\\
        \small $^{a}$School of Mathematics and Statistics, University of Glasgow, Glasgow, United Kingdom \\
        \small $^{b}$Dipartimento di Matematica e Fisica {\it Ennio De Giorgi},  Università del Salento\\ 
        \small $^c$  INFN, Sezione di Lecce, via Arnesano I-73100 Lecce, Italy
         \\\\
%
%
%
}
\date{\small{ \today}}

\maketitle
\begin{abstract}
We analyse a  generalised Fokker-Planck equation   by making essential use of its linearisability through a Cole-Hopf transformation.  We determine   solutions of travelling wave  and multi-kink type by resorting to a geometric construction in the regime of small viscosity. The resulting asymptotic solutions are time-dependent Heaviside step functions representing classical (viscous) shock waves.
As a result, line segments in the space of independent variables arise as resonance conditions of exponentials and represent  shock trajectories.
We then discuss fusion and fission dynamics exhibited by the multi-kinks  by drawing parallels in terms of shock collisions and scattering processes between particles, which preserve total mass and momentum. Finally, we  propose B\"acklund  transformations  and examine their action on the solutions to the equation under study.
 
 \end{abstract}


 \section{Introduction}\label{sec:intro}
 
 Models expressed by conservation laws and nonlinear partial differential equations (PDEs) play a crucial role in a number of fields, ranging from fluid dynamics \cite{bennet} and 
condensed matter physics \cite{epshteyn1} to population dynamics \cite{murray} and social sciences \cite{furioli}, and to chemical systems \cite{Field}.  Paradigmatic examples are provided by pivotal nonlinear PDEs with diffusion that have attracted huge attention in past decades, such as  
the Bateman-Burgers equation $ \partial_t v+v  \partial_x v=D \partial_{xx}v$,  or  more general nonlinear Fokker-Planck models 
$\partial_t v+D \partial_{xx}v=\partial_x [F(v) v]$,  being $D$ the viscosity/diffusion coefficient. Indeed,  these 
models very naturally arise while dealing with the mean field approximation of many-body systems \cite{frank}, 
incorporating various significant phenomena such as phase transitions, travelling-wave solutions, shock waves, and  anomalous diffusion. 

In this paper, we address the study of a nonlinear partial differential equation that can be considered a generalisation of the aforementioned models, that is
\bea & &\partial_{t}v(x,t) +\partial_{x} \lgr \frac{ c_2
	v(x,t)^2+c_4 \sigma^2 +\eta  \sigma  \left[c_1 \partial_{t}v (x,t)+c_2 \partial_{x}v(x,t)\right]}{ { c_1} v(x,t)+ { c_3} \sigma} \rgr = 0 \,,
 \label{GLMeq} \eea
 where $v=v(x,t)$ is a real function of the two real independent variables $x$ and $t$, $\eta$ is thought to be a small positive parameter, and  $\sigma$ and the $c_j$’s are real constants. This equation has been introduced recently in \cite{1} in connection with the problem of describing fluid systems of volume $v$ whose behaviour at pressure $P = x/t$ and temperature $T = 1/t$ deviates from the van der Waals equations of state.
However, its potential applicability is far more general, in principle. Indeed, Equation \eqref{GLMeq} is basically a conservation law, albeit not in a familiar evolutionary form  
  $\partial_{t}v=F(x,t, v,\partial_{x}v,\partial_{xx}v,\dots)$,   for which many general  results are available in the literature  (see e.g.  \cite{vinogradov, olver, sokolov, abramenko, lagno}).  As a matter of fact, \eqref{GLMeq}  belongs to the family of  continuity equations of the form $\p_t v=\p_x  J(v,\partial_x v,\partial_t v)$. Compared to standard nonlinear Fokker-Planck models, more drift and diffusion mechanisms are accounted.  Expressly for Eq. \eqref{GLMeq}, one recovers indeed the current $J=F(v) [g(v)+\eta\sigma(c_1\p_t v+c_2 \p_x v)]$, with $F(v)=(c_1 v+\sigma c_3)^{-1}$ and $g(v)=c_2v^2+c_4\sigma^2$. Accordingly, Eq. \eqref{GLMeq} can be seen as a generalised Fokker-Planck equation whose structure is of the type 
\beq
  \p_t v+D(v) \p_{xx}v+\tilde{D}(v) \p_{x,t} v =\p_x [F(v)g(v)] -\eta \sigma c_1 F(v)^2 (c_1 \p_t v+c_2 \p_x v) \p_x v \,\, , 
  \eeq
 where $D(v)=-\eta \sigma c_2 F(v)$, $\tilde{D}(v)=-\eta \sigma c_1 F(v)=c_1 D(v) /c_2$. Remark that for $c_1=0$ an equation of the Bateman-Burgers type follows, which is a fundamental nonlinear  PDE renowned for its complete integrability,  due to its remarkable property of being linearisable by means of a Cole-Hopf transformation (see e.g.  \cite{olver} and Refs. therein). Equation \eqref{GLMeq} may be also seen as a sort of variation of the Bateman-Burgers equation  similar to the Benjamin-Bona-Mohoney model   \cite{benjamin}   in the dissipative case, or  to  other analogous PDEs.

There are several ways to approach the study of solutions to  \eqref{GLMeq}, and their classification based on their properties. 
For instance, one may consider what happens in the \emph{inviscid} case $\eta\to 0$ as in \cite{1,GLMM},  where multivalued solutions replaced by  shock-type discontinuities have been evidenced \cite{1}. Moreover, one may investigate the symmetries of the equation within a Lie-group  framework to characterise analytically the similarity solutions on symmetry group orbits, as in \cite{GLM}, where attention has been also paid on pole dynamics by resorting to standard procedures. First in  \cite{1,GLMM}, then more recently in  \cite{GLM}, several special classes of solutions have been therefore considered that show rich dynamics and phenomenology.  Nevertheless, additional questions can still be raised  paving the way for achieving a clearer and broader perspective.
	 
In the present work, we continue the analysis of Eq. \eqref{GLMeq}  unveiling novel features of its solutions and  its integrability structure. 
To do so, we will pay our attention to the key-property of Eq.~ \eqref{GLMeq}, that is its linearisability via a Cole-Hopf  type transformation \cite{1}. Such property, although  being strongly motivating in the original derivation of the equation, has indeed received limited attention thus far in respect to potential implications and applications.  
From the   integrable systems perspective, this means that Eq. \eqref{GLMeq} is an analogous of the  Bateman-Burgers equation, suggesting that the problem   \eqref{GLMeq} may exhibit a plethora of integrability properties  \cite{LRB}.  Thus, it is certainly
  helpful to   assess Eq. \eqref{GLMeq} against techniques commonly employed to determine 
 special classes of solutions to the Bateman-Burgers or other diffusive equations. One of such techniques is the Hirota bilinear method, which plays  an important role in solving nonlinear integrable PDEs, see e.g. \cite{Hirota3,Hietarinta1}, unveiling often the existence of  multi-soliton solutions.  The existence of B\"acklund transformations for Eq. \eqref{GLMeq} may be devised too, thus enhancing the analysis of its solutions and integrability properties.
By making use of the Hirota  method and B\"acklund transformations, we will   identify soliton solutions  of multi-kink type to Eq.~\eqref{GLMeq}, which are underpinned by the implementation of suitable nonlinear superposition formulae.
 
The paper is organised as follows. In Section 2,  we discuss some features of Equation \eqref{GLMeq},
by identifying two distinguished   scalings, depending on conditions  for the coefficients $c_j$. 
In Section 3, we derive travelling wave solutions for the two  equations  implied by the scalings, and highlight their dynamical features. In Section 4, we tackle the problem of finding multi-kink solutions and carefully detail preeminent questions implied in their dynamics. We show the formation of classical (viscous) shocks, whose trajectories are provided by resonance conditions, and identify    mass and momentum conservation laws. This allows us to describe the interaction among shocks in terms of scattering among particles, establishing an intriguing shock-particle duality.  Section 5 is devoted to  B\"acklund transformations (BTs). We prospectively  devise   one- and two-parameters BTs and analyse their action on an initial solution. We show explicitly how BTs for Eq. \eqref{GLMeq} generate  new travelling components, which add  to  seed solution of the form derived in Section 2.  Section 6 is devoted to discussion and conclusions.

\section{Symmetries and rescalings} 

Equation  \eqref{GLMeq} contains six phenomenological parameters whose individual values have to be fixed pertaining the specific problem to be addressed. 
In this section, we perform scalings on variables and coefficients to simplify the analysis of Eq.~ \eqref{GLMeq} and better comprehend special cases and limits. 
For instance, we can take into account the two classical limits $\text{i)}$ and  $\text{ii)}$ below. 

\begin{enumerate}[ i)]
\item The inviscid (dispersionless) limit $\eta \to 0$ and $ c_2 \, c_3\, \sigma \neq 0$, from which, 
along  with the rescaling $t \to t = \frac{c_2 }{c_3 \sigma }\, t$,   one obtains a Riemann-Hopf type equation 
\beq {\partial_t\, v(x,t)}+\frac{\sigma  c_3}{c_2}  \partial_x\,\left[ \frac{\sigma
   ^2 c_4+c_2 v(x,t)^2}{\sigma  c_3+c_1 v(x,t)} \right]= 0 \,\,   .
   \label{Inviscid} \eeq
    Equation \eqref{Inviscid}  admits the general solution implicitly expressed by  the {\em hodograph transformation} 
\beq x-t\; {\partial_v} \left[
\frac{c_2 v(x,t)^2+c_4 \sigma ^2}{c_1   v(x,t)+c_3 \sigma } \right]= \psi \left[ v(x,t) \right] \,\, , \label{hodograph}\eeq
   where the arbitrary function $\psi$ determines the initial data for the implicit solution $ v\lf \xi\lf v \rg \rg = v\lf x - c\lf v \rg t \rg  $, and then its singularities, which propagates at the characteristic velocity
   $$ c\lf v \rg = \frac{c_2}{c_1}-\frac{ \sigma ^2 \left(c_4 c_1^2+c_2 c_3^2\right)}{c_1 \left(c_3
 \,  \sigma +c_1 v \right)^2}\,\, . $$   
   Remarkably, the non trivial dependency on $v$  of $c( v )$ is determined by the structural parameter of the model 
 \beq   \Delta := c_4  c_1^2 +c_2 c_3^2 \,\, . \label{Delta} \eeq  
In fact, when $\Delta$ vanishes,  one merely has that the initial datum $\psi^{-1}$  propagates at the constant speed $c = \frac{c_2}{c_1}$. This suggest that in several circumstances it is useful to perform the Galilei transformation \beq   \tilde{X} =  x-\frac{c_2}{c_1} t \,\, . \label{Galilei} \eeq  

\item   The  limit $c_1 \to 0$ and $ c_2 \, c_3\, \sigma \neq 0$ which, combined with  the time scaling $t \to t = \frac{2 c_2 }{c_3 \sigma } t$ 
provides the Bateman-Burgers equation $\p_t v\lf x, t\rg+v\lf x, t\rg \p_x v\lf x, t\rg +D  \p_{x x}v\lf x, t\rg = 0$  with $D  = \half \eta \sigma $  representing the effective viscosity.  
  
\end{enumerate}  
The previous two limits commute, leading to the Riemann-Hopf equation $\p_t \,v(x,t) + {\partial_x} \left[  v(x,t)^2\right] = 0$.  Moreover, it is well-known that the integrability properties of the two type of equations  relies on the existence of infinitely many commuting flows, or symmetries. In the inviscid case,  Eq.  \eqref{Inviscid} commutes with any flows described by similar equations of the type $\p_{t'} v\lf x, t, t'\rg + \p_x F\lfq v\lf x, t, t'\rg \rgq = 0 $  for any differentiable function $F$.  On the other hand, also the Bateman-Burgers equation admits  an infinite  numerable  set of commuting flows,  recursively defined by \cite{olver2}
 \beq \p_{t_j} v = \p_x P_j \,\, , \eeq
where
\beq
   P_j =
     P_{-1} \left[ \frac{v(x, t_0, t_1,t_2, \dots , t_j , \dots )}{ 2D }+  {\partial_x} \right] \,P_{j-1} \,\, ,
     \qquad \quad P_{-1} = \sqrt[3]{2} D ^{2/3} \,\,  ,
 \eeq
and $j=0,1,2,\dots$.    The Bateman-Burgers equation    corresponds to $j=1$.

 For what  follows, it could be of some interest the limit $c_2 \to 0$ of the Eq. \eqref{GLMeq}. It yields 
   \beq {\partial_t\, v(x,t)} + {\partial_x} \left[ \frac{\sigma ^2 c_4+\eta 
   \sigma  c_1 \, {\partial_t\,  v(x,t)}}{\sigma  c_3+c_1 v(x,t)} \right] = 0 \,\, , \label{preScaled}\eeq
   which in the inviscid limit provides again an equation of  Riemann-Hopf type as in the Bateman-Burger case, but with a singular current. 
   In this sense, the limit $c_2 \to 0$ provides a class of equations qualitatively distinct from the Bateman-Burgers one. On the other hand, it is exactly for these models, obtained by the double limit $c_2 \to 0 , \; \eta \to 0$,  that one can interpret certain classes of solutions as equation of state for the van der Waals gases \cite{1}, as can be easily seen for particular  choices of the function $\psi$ in \eqref{hodograph}.

For the general Eq. \eqref{GLMeq} one would like to verify whether it shares the beautiful set of properties  possessed by the limit cases  i) and ii) above. In doing this, by introduction of  $\Delta$  in \eqref{Delta}  one already noticed that the whole space of the model parameters $\lgr c_1, c_2, c_3, c_4 \rgr$ can be decomposed  in   two regions  according to the two  cases 
\beq
\Delta \neq 0 \qquad \text{and} \qquad \Delta=0 \,\, . 
\label{subDelta}
\eeq
Such a distinction is supported by the results concerning the  point symmetries analysis of the Equation \eqref{GLMeq}, signalling quite a relevant difference out  the two  cases \cite{GLMM}: there exists a finite algebra in the first case  and an infinite dimensional symmetry algebra  in the second case.  To shed some further light on this point,
  we observe here  that the  physical essence  of the specific constraint   $\Delta=0$ relies on the long-wave approximation methods for nonlinear dispersive PDEs in the {\em far-field} limit \cite{kahawara,jeffrey}.   Indeed, if we split Equation \eqref{GLMeq} in the form $\hat{\cO}_L v+\hat{\cO}_{NL}v=0$ to explicitly separate its  linear and nonlinear parts, where

\newcommand{\cev}[1]{\reflectbox{\ensuremath{\vec{\reflectbox{\ensuremath{#1}}}}}}
\begin{align}
&\hat{\cO}_L v= \sigma \left[ c_3^2 \partial_t-c_1c_4 \partial_x+\eta c_3 \partial_x(c_1\partial_{t} +c_2\partial_{x})  \right] \, v\,\,  ,
\label{eq: O L}\\
&\hat{\cO}_{NL}v= v\left[\eta c_1(\partial_x-\cev{\partial}_x )+2c_3 \right] (c_1 \partial_t v+c_2 \partial_x v) \,\, ,  
\end{align}
and  we look for real solutions of just the dispersive part $\hat{\cO}_Lv=0$  having the form $v=v^{(0)} e^{kx - \omega_{\hat{\cO}_L} (k) t}$, we  find 
 the dispersion relation 
\beq
\omega_{\hat{\cO}_L}(k) =\frac{\eta c_2c_3 k^2-c_1c_4 k}{\eta c_1 c_3 k+c_3^2} \,\, .
\label{eq: dispersion OL}
\eeq
Since the \emph{group velocity}  is expressed by 
\[
\p_k^2 \omega_{\hat{\cO}_L}(k)=\frac{ 2 \eta \, \Delta}{(c_1 \eta  k+c_3)^{3}} \,\, ,
\]  
we therefore become aware that  when $\Delta =0$ the problem is no longer {\em purely dispersive}, the dispersion relation being 
just linear: $\omega_{\hat{\cO}_L}(k)=\frac{c_2}{ c_1} k$. 
When $\Delta\neq0$  a non trivial dependence of the group velocity on $k$ results instead.  Of course, when the \emph{phase}  $k x - \omega_{\hat{\cO}_L} (k) t \gtrsim 1$ the nonlinear contributions described by $\hat{\cO}_{NL}  v$ become relevant for the description of the solutions and modulation instability has to be considered.

The above characterisation \eqref{subDelta} allows to perform  distinct coordinate transformations on Eq.  \eqref{GLMeq}, which  can be reformulated in suitable forms  for further analysis. Furthermore, let us notice that both the previous limits i) and  ii) can be performed  in both the coefficients manifolds determined by \eqref{subDelta}.  This approach allows for a deeper understanding of the solutions and properties associated with each case. 

\subsection{Case 1: $\Delta \neq 0$}

 When $\Delta  \neq 0$ and $c_1,c_3,\sigma \neq0$, it is useful to perform the change  into the dimensionless  variables 
	\beq X =
	 \frac{\tilde{X}}{\Delta}\,\,  ,  \quad  \quad T = \frac{t}{c_1 c_3^2} \,\, ,  \quad 
	  \quad u\lf X, T \rg=  \frac{c_1}{ c_3 \sigma }  v + 1 \,\, , 
	 \label{CRvisco}
	 \eeq 
so that Eq. \eqref{GLMeq} can be rewritten  in the form 
	\beq \left[ 1 - \epsilon \, \p_X \lf \frac{ \bullet}{u}\rg \right] u_T = - \p_X   \frac{1}{u} \,\, ,  \label{eqViscoPert}\eeq 
with	 $\epsilon =\frac{c_1 \eta }{c_3 \Delta} $ representing the dimensionless viscosity parameter and $\left[\partial_X \lf \frac{ \bullet}{u}\rg \right] u_T =\partial_X \lf \frac{ u_T}{u}\rg $.  The  symmetry algebra for this differential problem is generated by the vector fields 
	                                  \beq
					 W_1=
 					\frac{\partial}{\partial X} \,\, , \qquad 
 					W_2=
 					\frac{\partial}{\partial T} 
 					\,\,, \qquad W_3= T \frac{\partial}{\partial T} -  X \frac{\partial}{\partial X} +
 					\, u \frac{\partial}{\partial u} \,\, .
 					\label{W1 W2 W3}
 					\eeq		
The meaning of the generators  is self-evident,   $W_1$ and $W_2$  are associated with rigid translations in the  $X$ and $T$  directions, while  $W_3$ is a dilation 
(for details about the corresponding symmetry reductions see \cite{GLMM, GLM}).  In summary, it is a 3-dimensional solvable algebra, which can be easily studied and integrated, in order to generate the Lie  group of symmetries, allowing to map solutions of \eqref{GLMeq} among themselves.
Besides,    from \eqref{eqViscoPert}  it is seen that the equation admits the further discrete extended $\cP \cT$ symmetry 
\begin{equation}
\label{eq:PT symmetry}
\widetilde{\mathcal{P T}}:  	(T,X,u) \to -(T,X,u) \, \,.
\end{equation}
When the the viscosity parameter $\epsilon$  is small, the equation \eqref{eqViscoPert} can be considered as a singular perturbative problem, which can be dealt with various methods \cite{Nayfeh,Bender,OMalley}. But, out of  the  kernel  of $ \lf 1 - \epsilon \, \p_X \frac{\bullet}{u} \rg$, its  inverse operator can be developed in power series,  obtaining  the formal evolutionary expression  
\beq  \partial_T u= - \p_X \lfq \sum_{n = 0}^\infty \; \epsilon^n \; \lf  \frac{1}{u} \p_X \, \bullet\rg^n \rgq \frac{1}{u} \,\, , \label{evolform}\eeq 
which makes sense if the series in  r.h.s. is  convergent. 
	To this equation can be applied the classification procedure of the viscous conservation equation due to \cite{arsie}. To proceed in this direction, let us expand the first terms of Eq. \eqref{evolform}, namely 
	\beq  \partial_T u = - \p_X \lfq  \frac{1}{u }-\frac{\epsilon 
   \partial_X u}{u ^3}-\epsilon ^2 \left(-\frac{3
      ( \partial_X u)^2}{u ^5}+\frac{ \partial^2_X u  }{u ^4}\right)-\epsilon ^3 \left(\frac{15   (\partial_X u)^3}{u ^7}-\frac{10
     \partial_X u \,
    \partial^2_X u}{u ^6}+\frac{ \partial^3_X u}{u ^5}\right) + O\lf \epsilon^4 \rg\rgq \,\,.
    \label{eq:evolutive for u}
    \eeq
    Clearly,  all terms are derived uniquely by a well defined rule from the first one, leading to an equation  of the form  of a viscous conservation law 
\beq  \partial_T u  = \p_X \left\{ f\lf u \rg + \epsilon A(u)  \partial_X u + \epsilon \left[ B_1(u) \partial^2_X u  + B_2(u)  (\partial_X u)^2 \right] +O\lf \epsilon^2 \rg \right\} \,\, .\eeq 
The limit of \eqref{evolform} for $\epsilon\to 0$ is $ \partial_T u = - \p_X  u^{-1}$, which is different from that one for the Burgers, KdV and Camassa-Holm equations. 
However, performing the transformation $u=1/ \sqrt{z(X,T)}$ the dispersionless limit of the \eqref{evolform} can be written as the Riemann-Hopf equation $\p_T z=\p_X(z^2/2)$, common to all other above quoted equations. Furthermore, the so-called {\em central viscous invariant} can be calculated by the formula  $a( u) = \frac{2  A(u)}{ \p_u^2 f(u)}$, giving $a(u)=-1$ \cite{arsie}.
This quantity is invariant under general Miura transformations, mapping equations into others in the same class: a constant viscous central invariant is characteristic of  the Bateman-Burgers hierarchy and it determines uniquely all the higher order terms. This suggests the zeroth order term 
of a Miura transformation of the form  $ u =  \frac{1}{\sqrt{z\lf X, T\rg}}+ \sum_{j = 1}^\infty \epsilon^j F_j\lfq z, \partial_X z, \partial^2_X z, \dots , \partial^j _X z\rgq $, allowing to map  \eqref{evolform}  into the Bateman-Burgers equation.

Before   we proceed further in this direction, let us  introduce the new  set of dimensionless variables
\beq 
 X = \frac{c_3 }{c_1 \eta} \tilde{X} \,\, , \quad T = \frac{\Delta }{c_1^2 c_3   \eta } t \,\, , \quad 
 u\lf X, T \rg=  \frac{c_1}{ c_3 \sigma }  v + 1 \,\, ,
\label{TraslScal}
\eeq
which brings Eq.~\eqref{GLMeq} into the following parameter-free form
\beq 
 \partial_{T} u(X,T)  +     \partial_{X}             \left[\frac{1 +   \partial_{T}   u(X,T)   }{u(X,T)}   \right] = 0 \,\, .
\label{eqscaled} 
\eeq
Notice that Eq.~\ref{eqscaled} corresponds to Eq. \eqref{eqViscoPert} upon setting  $\epsilon =1$ and $u \to -u$. Moreover, Eq. \eqref{eqscaled} should be considered in the light of the comments to Equation \eqref{preScaled}.
 In other words,  under the given hypothesis 
the  Eq.  \eqref{GLMeq}  can be always  reduced to the special case \eqref{eqscaled}, corresponding to the \emph{universal } special model    $c_1 = \eta \sigma = c_4 \sigma^2 = 1$ and $c_2 , \, c_3 \to 0$. The inviscid limit being recovered in the  limit $\eta \to 0$ under the condition $\sigma \, c_3 \neq 0$ fixed, if it exists.  In such a case, any solution of Eq.  \eqref{eqscaled} holds also for $\eta \to 0$. It is now to be pointed out that we actually end up into an equation of the form \eqref{eqscaled}  even whenever $c_3=0$,  with $c_1 c_4 \neq0$.   In such a case, indeed, we can define the variables 
\beq
X=\frac{c_4}{c_1 \eta^2} \tilde{X} \,\, ,  \quad T=t\,\, , \quad v= \frac {c_4\sigma}{c_1\eta} u(X,T) \,\, , 
\eeq
to cast \eqref{GLMeq} again into \eqref{eqscaled}.  Taking   $c_3=c_4=0$,   Equation \eqref{GLMeq} can be solved by direct integration (see Eq. (10) in \cite{GLM}).  A similar conclusion can be drawn when $c_1=0$, when a Bateman-Burgers equation arises and, as said above,  the general solution can be given in terms of a linear heat-type equation. To summarise,   by relaxing the coefficients' condition $c_1c_3\neq0 $ we are concerned with two solvable cases, for which general solutions are known if the initial value problem is assigned. 
 
\begin{remark}
 {\em 
 By introducing the function
 \beq 
\rho = u + \partial_X \log u 
\,\, , 
\label{defrhoJ}
\eeq
Eq.  \eqref{eqscaled}  can be put into the conservation law form $\partial_T\, \rho + \partial_X\, \tilde{J} = 0$, with $\tilde{J} = \frac{1}{u}$. Equivalently,  
it can be seen as the system 
\bea
 \p_X  u=  u \lf  \rho -u  \rg  \,\, , \qquad  \p_T \rho =  -\partial_{X}\lf \frac{1}{u}    \rg \,\, .
\label{systurho}
\eea
This latter form can be useful in studying the evolution of an initial datum $u\lf X, T=0 \rg = u_0 \lf X\rg$.  In fact, owing to the definition \eqref{defrhoJ}, one has an initial datum also for   $\rho\lf X, T=0 \rg = \rho_0 \lf X\rg $.  Then,  by resorting to the second equation in \eqref{systurho}, one updates the function $\rho \lf X, \delta T \rg = \rho_0 \lf X\rg - 
\partial_{X}\lf \frac{1}{u_0 \lf X\rg}\rg \delta T$. But using this result into the first equation, one must assume $\p_X  u_0\lf X\rg \approx   u_0 \lf X\rg \lf \rho \lf X, \delta T \rg  - u\lf X, \delta T \rg   \rg$ modulo higher order infinitesimals. Hence, one gets the updated function   $u\lf X, \delta T \rg \approx (\partial_{X}u_{0}\lf X\rg)/u_0\lf X\rg - \rho \lf X, \delta T \rg   = u_0\lf X\rg - \p_X \lf \frac{1}{u_0\lf X \rg}\rg $. The procedure can be now repeated in order to obtain an approximated expression for    $u\lf X, 2 \delta T \rg$. It should be remarked here that the above expression for  $u\lf X, \delta T \rg$ is exactly what one should obtain at first order in $\delta T$ and for $\epsilon \to 0$ from Eq. \eqref{eq:evolutive for u},  confirming the consistency of the treatment. However, the stability and the convergence  of the procedure must be studied more attentively. This in view of the possible divergencies of the function $\rho$ in correspondence of the zeroes of $u$. 
On the other hand, when one deals with the first equation in \eqref{systurho} by the different approximation  
$\p_X  u \lf X, \frac{\Delta T}{2}\rg \approx   u_0 \lf X\rg \lf \rho \lf X, \delta T \rg  - u_0\lf X \rg   \rg$.  This relation allows us to the expansion
$u \lf X, \frac{\Delta T}{2}\rg \approx   u_0\lf X \rg + u_0 \lf X\rg \lf \rho \lf X, \delta T \rg  - u_0\lf X \rg   \rg \Delta X +O\lf \Delta X^2 \rg$, which can be used to compute $\rho \lf X, \frac{3 \Delta T}{2}\rg $ from the second equation in  \eqref{systurho} and so on. Also in this case a more detailed stability analysis is required.   
}
\end{remark}

\subsection{Case 2: $\Delta = 0$}

When $\Delta  =0$, Eq.  \eqref{GLMeq} admits an infinite point symmetry sub-algebra  \cite{GLMM}, implying its solvability. By  simply performing  the Galilei change of reference \eqref{Galilei} into variables  variables $( \tilde{X}, T = t, u = \frac{c_1}{c_3 \sigma} v + 1)$, the symmetry vector-fields take the form  
\beq
W_{f_1, f_2} =  f_2( \tilde{X} ) \,\p_{\tilde{X}} +  f_1\lf T\rg   \p_T - \frac{\sigma c_3}{c_1}   f_2' ( \tilde{X} ) \,u\,  \p_u 
\,\,  , 
\label{generators infinite algebra}
\eeq 
$f_1, \;f_2$ being  arbitrary functions of their argument. Eq.  \eqref{generators infinite algebra} is 
evidently  the direct sum of the Cartan’s infinite-dimensional simple Lie algebra \cite{Cartan}  of all  real smooth vector fields on $T$ with   another isomorphic one,  acting on the 
$\tilde{X} - u$ plane by arbitrary differentiable local transformations in the $\tilde{X}$ variable, but at most of first degree in the variable  $u$.  A basis of the first subalgebra is given by the vector-fields $\tau_n =  \frac{c_1}{c_2} T^n \, \p_T $, the commutation relation being $\lfq \tau_m, \, \tau_n \rgq =  \lf n-m \rg \frac{c_1}{c_2}   \tau_{m+n-1}$. A basis of the other subalgebra is given by the generators   $\sigma_n = \tilde{X}^n \p_{\tilde{X}} - n  \sigma \frac{c_3}{c_1}  u \tilde{X}^{n-1} \, \p_v$,with commutation relation $\lfq \sigma_m, \, \sigma_n \rgq = \lf n-m \rg  \sigma_{m+n-1}$. This symmetry structure suggests that the equation is solvable by a suitable transformation, in which the $T$ variable plays a mere auxiliary role. In fact, assumed  that $c_1 c_3 \neq 0$, one can use the dimensionless variables $X$ and $u$, scaled with respect the viscosity parameter  $\eta\neq 0$, namely 
\beq 
X= \frac{c_3 }{c_1  \eta } \tilde{X} \,\, , \quad T = t  \,\, ,  \quad v\lf x, t \rg = \frac{c_3 \sigma }{c_1}  \left[ u\left( X, T \right)-1\right] \,\, , \label{CRefScaling}
\eeq
leading   to  the parameter--free equation
   \beq 
   \partial_{T}
 \left[u(X , T)+\frac{\partial_{X}u( X,  T)}{u( X,  T)} \right] = 0 \,\, . 
   \label{eqscaled2}\eeq 
 By introducing an integration arbitrary function $g( X)$, the above equation takes   the Riccati form 
   \beq
\partial_{X} u( X,  T)= - u( X, T)^2+g( X) u( X, T) \,\, .
\eeq 
  Then, setting $ g(X)=[\ln B'(X) ]'=\frac{B''}{ B' } $, where  $'=\frac{d}{d X}$, one is able to find the general solution
    \beq 
    u(X,T)=\partial_{X}  \ln [A(T)+B(X)]   
        \label{sol Delta 0}
    \eeq
for arbitrary functions  $A(T)$ and $B(X)$, determined by the imposed  boundary valued problem  to Eq.  \eqref{GLMeq}.
The transformation \eqref{CRefScaling}  holds for any $\eta\neq 0$, so that the inviscid limit $\eta\to 0$ is captured when $|X| \to \infty$. 
Since when $\Delta=0$ Eq.~\eqref{GLMeq} is exactly solvable through the mere assignment of an initial value problem, 
in what follows we will  investigate mostly properties underlying the case  $\Delta\neq0$.

\subsection{Cole-Hopf Transformation  for Equations~\eqref{eqscaled} and \eqref{eqscaled2}}

\noindent The exact solvability of Eq. \eqref{GLMeq} follows from the linearisation of Equations \eqref{eqscaled} and \eqref{eqscaled2} via a Cole-Hopf type transformation, namely
\beq
u=\partial_X \log \phi(X,T) \, .
\label{CH X}
\eeq
 In both  cases, the result can be also recognised as the realisation of an elementary Hirota bilinear $D$ operator equation constraining two  connected simple linear operator problems  for the  function $\phi(X,T)$.

\subsubsection{ Case 1: Equation \eqref{eqscaled}}

\noindent \noindent 
In this case,  we can introduce two operators 
		
		\beq
		 \hat{\Lambda}_1 := \partial_X \quad ,\quad  \hat{\Lambda}_2 :=1+  \partial_X \partial_T=1+ \partial_T \hat{\Lambda}_1 \, \,  ,
		 \label{Lambda1 Lambda2} 
		 \eeq
		such that Equation \eqref{eqscaled} is mapped via \eqref{CH X} to the following bilinear problem:
			\begin{align}
			\notag
					\p_T u+ \partial_{X}  \left( \frac{1 +   
						\p_Tu}{u} \right) &=\partial_X \Big\{ \left[ \hat{\Lambda}_1 \phi(X,T)\right]^{-1}  \left[ \hat{\Lambda}_2 \phi(X,T)\right] \Big\} \\
			&= \left[ \hat{\Lambda}_1 \phi(X,T)\right]^{-2}  \mathcal{D}_X \left[ \hat{\Lambda}_2 \phi(X,T)  \cdot \hat{\Lambda}_1 \phi(X,T) \right]
			\nonumber \\
			&=0\,\, ,
			\label{eq Hirota scaled}
		\end{align} 
where one notices that $u\neq 0$ implies $\hat{\Lambda}_1 \phi(X,T)\neq0$, and $\mathcal{D}_X$ stands for the standard Hirota derivative w.r.t. variable $X$, 
\beq 
 \mathcal{D}_X\lf \phi_1\cdot \phi_2 \rg= \lf \p_X -\p_{X'}\rg \, \phi_1\lf X, T\rg  \, \phi_2\lf X', T'\rg  \Big|_{X' = X, T' = T} \,\, .
\label{HirotaD X}
\eeq		 
From \eqref{eq Hirota scaled} one concludes that the equation is equivalent to the linear problem $ [ \hat{\Lambda}_2+ Z_1(T) \hat{\Lambda}_1]  \phi(X,T)=0$ where 
$Z_1(T) $ is an arbitrary function, and that this equation can be forthwith converted to just  
\beq 
\hat{\Lambda}_2 \phi(X,T):= \p_{XT} \phi +\phi = 0 \,\,  
\label{KGeq} 
\eeq
 by  a gauge transformation $\phi(X,T) \to e^{-\int^T Z_1(T')dT'} \phi(X,T) $.

 \begin{remark}
 {\em 
Equation \eqref{KGeq}  can  be derived from the system   \bea 
\p_X \phi  = u \, \phi \,\, , 
\qquad 
\p_T \phi =  - \frac{\left(1+\p_T u\right) }{u } \phi \,\, , \label{syslinprobl}
\eea
whose compatibility condition,  $\phi _{X T} = \phi _{T X}$,  provides Eq.  \eqref{eqscaled}.  These relations allow to discuss the initial boundary value  problem for \eqref{eqscaled} by solving  the system \eqref{syslinprobl} for $\phi$, for which one needs to provide an initial datum $u\lf X, 0 \rg = u_0\lf X\rg$, but also to assign a  function, say  at $X = 0$, $u_T\lf 0, T \rg = u_1\lf T\rg$. 
Both functions  provide, through Eqs. \eqref{syslinprobl},   the initial and boundary conditions $\phi\lf X, 0\rg$ and $\phi\lf 0, T\rg$
 for the second order equation \eqref{KGeq}.  Finally, using again the first  equation in  \eqref{syslinprobl}, one obtains the desired solution to the nonlinear equation \eqref{eqscaled}.
}
\end{remark}

\begin{remark}
 {\em 
By introducing $\Phi = (\Phi_1,\Phi_2)^\mathbf{\top}$, Equation \eqref{systurho}
can be written equivalently as the integrability condition ${\bf{T}}_X-{\bf{X}}_T+[{\bf{T}},{\bf{X}}]=0$ for the overdetermined linear matrix problem
	\beq  
	\label{eq:AKNS pair}
	\p_X \Phi = {\bf{X}} \, \Phi, \qquad  \p_T  \Phi ={{\bf{T}}} \, \Phi \,\, , \eeq
	where the two matrix operators are
			\beq
		{\bf{X}}= \left(
	\begin{array}{cc}
		u & 0 \\
		0 & \rho  \\
	\end{array}
	\right) \quad, \quad {\bf{T}}= \left(
	\begin{array}{cc}
		- \frac{1+\p_T u }{u} & 0 \\
		\alpha \, u & -\frac{1}{u} \\
	\end{array}
	\right) \,\, . 
	\label{eq: Lax X T}
	\eeq
The ${\bf T}$ matrix depends on an arbitrary parameter  $\alpha \in\mathbb{ R}/\lgr 0\rgr$, which may play a role analogous to the spectral parameter in the AKNS formalism 
\cite{ablowitz}. 
}

\end{remark}

\subsubsection{Case 2:  Equation \eqref{eqscaled2}}

\noindent Once we focus to the case $\Delta=0$, for which  Eq.~\eqref{GLMeq} is brought to the simple form  Eq.~\eqref{eqscaled2}
upon the change of variables \eqref{CRefScaling}, the Cole-Hopf type transformation \eqref{CH X} gives
\begin{align}
	\notag
   \partial_{T}
  \left( u + \frac{   
		\p_X u}{u} \right) &=\partial_T \Big\{ \left[ \hat{\Lambda}_1 \phi(X,T)\right]^{-1}  \left[ \hat{\Lambda}_3 \phi(X,T)\right] \Big\} \\
	&= \left[ \hat{\Lambda}_1 \phi(X,T)\right]^{-2}   \mathcal{D}_T \left[ \hat{\Lambda}_3 \phi(X,T) \cdot \hat{\Lambda}_1 \phi(X,T) \right]=0\,,
\end{align}
where we have introduced the operator  $\hat{\Lambda}_3:= \partial_X^{2} =  \left(\hat{\Lambda}_1\right)^2$  and the Hirota bilinear operator $\mathcal{D}_T$ as in  \eqref{HirotaD X},
 but clearly with 
$\p_X-\p_{X'} \; \to \;\p_T-\p_{T'}$.  The nonlinear problem \eqref{eqscaled2} is therefore mapped to a linear one specified by 
$\left[ \hat{\Lambda}_3 + Z_0(X)  \hat{\Lambda}_1 \right] \phi(X,T) =
\left[ \hat{\Lambda}_1 + Z_0(X)   \right] \hat{\Lambda}_1 \phi(X,T)=0$, or more  explicitely  $[\partial_X^2+Z_0(X) \partial_X ] \phi(X,T)=0$, whose solutions read 
\begin{equation}
\phi(X,T)= \phi_0(T)+\phi_1(T) \int^{X'} e^{-\int^{X'}Z_0(X'') dX''} dX' \,\ , 
\end{equation}
where $Z_0(X)$ is an arbitrary function of its argument.  Consistently,  solutions $u$ of the form \eqref{sol Delta 0} ensue in the end, the 
case   $Z_0(X)=0$ leading to Laplace's equation  for the auxiliary function $\phi(X,T)$ and rational solutions $u=[u_0(T) +X]^{-1}$, with $u_0(T)$  being an arbitrary function.

\section{Travelling waves }
\label{section travelling}

In this Section, we look for travelling wave solutions to the rescaled Eqs. \eqref{eqscaled} and \eqref{eqscaled2}, that is 
 solutions of the form $u(X,T)=U(\xi)$, where  $\xi=\lf  X - c\, T\rg$ is the travelling wave coordinate and  $c\in \mathbb{R}\setminus\{0\}$ is the constant velocity. This analysis will allow us to identify single-kink solutions, providing the basis for the analysis of multi-kink solutions
 in Section  4.   Moreover, single-kink solutions  will provide  a convenient class of seed solutions for the application of B\"acklund transformations, as presented in Section {\ref{backlund}}.

\subsection{Travelling wave solutions to Equation \eqref{eqscaled}}
\label{sec:TW sol eq scaled}
Travelling wave solutions to Eq. \eqref{eqscaled}, are solutions of the form 
\begin{equation}
	\label{eq:TW ansatz}
	u(X,T)=U(\xi) \,, \quad \xi=X-c\,T\,,
\end{equation}
where $\xi$ is an invariant coordinate under the  conjugacy class of  1-dimensional symmetry subalgebra $W_c=c W_1+W_2$, with  $c$ denoting the wave speed. All the elements of such a class are conjugated by the $Ad_{W_3}$ inner automorphisms, changing the velocity and the scaling $U$. The ansatz \eqref{eq:TW ansatz}  gives the following nonlinear ODE for the wave profile $U$,
\begin{equation}
	\label{eq:ODE scaled}
	\left[ -c \, U(\xi) + \frac{1}{U(\xi)} - \frac{c}{U(\xi)} U'(\xi)   \right]'=0 \,,
\end{equation}
where $'\equiv\frac{d}{d \xi}$. Integrating Eq. \eqref{eq:ODE scaled}, gives the following first order ODE of Riccati type,
\begin{equation}
	\label{eq:ODE Riccati eqscaled}
	U'(\xi) + (U-K_1)^2 - K_1^2 -\frac{1}{c}=0\,,
\end{equation}
where $K_1\in \mathbb{R}$ is an integration constant. Real solutions to Eq.~\eqref{eq:ODE Riccati eqscaled} change drastically depending on the sign of the constant term $K_1^2 +\frac{1}{c}$.

If $K_1^2 +\frac{1}{c}\geq 0$, solutions to Eq.~\eqref{eq:ODE Riccati eqscaled} are  given by\footnote{We should notice here that the $X$-component of the gradient of the single-kink solution is given by 
	\[ \partial_X u(X,T)= \left( K_1^2+\frac{1}{c} \right) \text{sech}^2 \left[
	\sqrt{K_1^2+\frac{1}{c}} \; (X-c \, T+K_2) \right] \,,\]
	hence resembling the shape of the KdV solitary wave. However, one should notice that the amplitude, in this case, is inversely proportional to the velocity of propagation. This implies that taller solitons move slower than shorter ones.  }
	\beq 
u\lf X, T \rg = K_1 +   \sqrt{K_1^2+\frac{1}{c}} \tanh \left[
\sqrt{K_1^2+\frac{1}{c}} \; (X-c \, T+K_2) \right] ,  \quad K_1, K_2 \in \mathbb{R}. \label{eq:TW scaled}
\eeq
Solutions of the form~\eqref{eq:TW scaled} represent single-kink which uniformly move left-to-right if $c>0$ or right-to-left if $c<-1/K_1^2$. Notice that the special case $c=-1/K_1^2$ returns the constant solution $u=K_1$. The parameter $ K_2 $ determines the  position of the kink  at $t=0$, while the parameter  $K_1$ characterises, together with the wave speed, its amplitude. Indeed, we have 
\[ \lim_{X \to \pm \infty} u(X,T) =K_1 \pm  \sqrt{K_1^2+\frac{1}{c}}\,,\]
which implies  a global step of  amplitude
\beq
{\cal A}(K_1,c):= \Big| \!  \lim_{X \to \infty} u(X,T) - \lim_{X \to - \infty} u(X,T)\Big| = 2\sqrt{K_1^2+\frac{1}{c}} \,\, .
\eeq

\begin{remark}
{\em 
According to the solution in Eq.\eqref{eq:TW scaled}, the size of the  region of high-gradient  is  $\delta X \sim {\cal A} ^{-1}(K_1,c)$. It is worth to stress that  the transformation \eqref{TraslScal} implies that, in the original coordinates $x$, $t$ and $v$, the region of high-gradient is proportional to $\eta$. }
\end{remark}

If  $K_1^2 +\frac{1}{c}<0$,  by analytic continuation of \eqref{eq:TW scaled}, one obtains the periodic singular solutions to Eq. \eqref{eq:ODE Riccati eqscaled} 
\begin{equation}  
	\label{eq:TW tangent}
	u\lf X, T \rg = K_1 -   \sqrt{-\left(K_1^2+\frac{1}{c}\right)} \, \, \tan \left[
	\sqrt{-\left(K_1^2+\frac{1}{c}\right)} \; (X-c \, T+k_2) \right] \,\,,  \quad K_1, K_2 \in \mathbb{R}.
\end{equation} 
Notice that solution~\eqref{eq:TW tangent} has infinitely many countable uniformly moving singularities at 
\[
X_n (T)= c \,T+ \frac{\pi}{2} (2n+1)\,\,,\quad n\in\mathbb{Z}\,\, .\]
The behaviour  of solutions~\eqref{eq:TW scaled} and \eqref{eq:TW tangent}  is displayed in Figure~\eqref{fig: travelling waves}.  As expected, the travelling wave profile translates uniformly along the $X$ coordinates left-to-right (if $c>0$) or right-to-left (if $c<0$).

\subsection{Travelling wave solutions to Equation \eqref{eqscaled2}}
\label{sec:TW sol eq scaled 2}
Similarly to the analysis carried in Section \eqref{sec:TW sol eq scaled}, looking for travelling wave solutions  $u(X,T)=U(\xi)$ to Eq. \eqref{eqscaled2} yields
\begin{equation}
	\label{eq:ODE scaled 2}
	\left[ U(\xi) + \frac{1}{U(\xi)}U'(\xi)   \right]'=0 \qquad \quad \lf '\equiv\frac{d}{d \xi}\rg\,,
\end{equation}
which can be straightforwardly integrated to give the  following first order ODE for $U$
\begin{equation}
	\label{eq:ODE eqscaled 2}
	U'(\xi) = (K_1-U)\,U \,\, , 
\end{equation}
where $K_1\in \mathbb{ R}$ is an integration constant.

\begin{figure}[h!]
	\centering 
	\includegraphics[scale=0.39,angle=0]{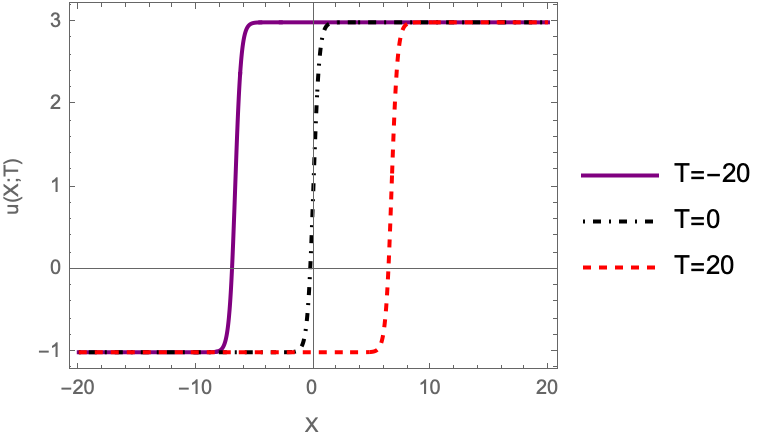}  
	\includegraphics[scale=0.39,angle=0]{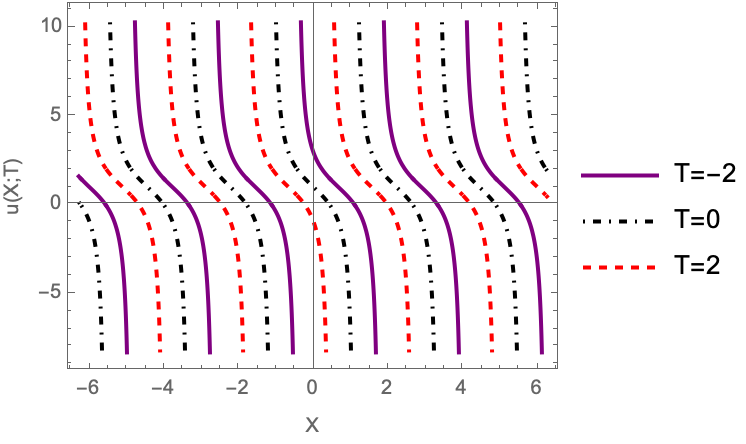} 
	\includegraphics[scale=0.39,angle=0]{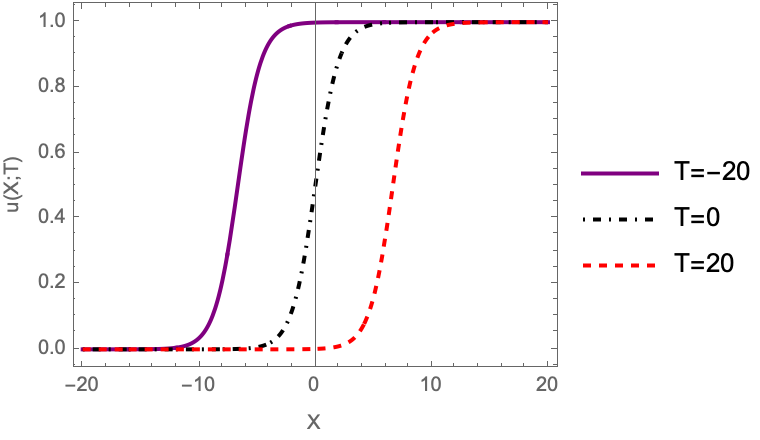}
	\caption{\small{Travelling wave solutions to Eqs.~\eqref{eqscaled} and  \eqref{eqscaled2}. \emph{(Left)} Solution~\eqref{eq:TW scaled} at different times for $K_1^2+1/c>0$. Parameters are $K_1=1$, $K_2=0$ and $c=1/3$. \emph{(Centre)} Solution \eqref{eq:TW tangent} at different times obtained for parameters satisfying $K_1^2+1/c<0$. In particular,  parameters are chosen to be  $K_1=1$, $K_2=0$ and $c=-1/3$.  \emph{(Right)} Solution~\eqref{eq:TW eq scaled 2} at different times. Parameters are $K_1=1$, $K_2=0$ and $c=1/3$. Notice that, at equal values of $K_1$ the kink in the case $\Delta \neq 0$  is steeper than the one  in the case $\Delta  =0$, due to 	dependence of the amplitude on the velocity.}}
	\label{fig: travelling waves}
\end{figure}

Equation \eqref{eq:ODE eqscaled 2} resembles the spatially homogeneous reduction of the Kolmogorov–Fisher reaction-diffusion model with logistic growth 
\cite{Du, peppo,Grindrod} for fixed linear growth–rate and carrying capacity $K_1$   \cite{murray}. 
Integrating Eq. \eqref{eq:ODE eqscaled 2}, we get the following travelling wave solution
\begin{equation}
	\label{eq:TW eq scaled 2}
	u(X,T)=\frac{K_1}{2} \left\{ 1+\tanh{\left[\frac{K_1}{2}(X-c\,T + K_2)  \right] } \right\}  \,\, , 
\end{equation}
where $K_2\in \mathbb{R}$ is an integration constant. The value of $K_2$ provides the initial location of the wavefront, meaning the location  at which the gradient is maximum (in modulus). Indeed, one notice that $u(K_2,0)=K_1/2$ and $\partial_{X}^2 u(K_2,0)=0$, with $K_1$ also assigning uniquely the asymptotic amplitude of the wave, 
\[ \mathcal{A} (K_1)= \,  \Big| \!  \lim_{X \to \infty} u(X,T) - \lim_{X \to - \infty}  u(X,T) \Big| =|K_1|\,. \]
Sigmoids \eqref{eq:TW eq scaled 2}   are displayed in the right panel of Figure~\eqref{fig: travelling waves}. Compared to the kink-type solutions obtained for Eq.~\eqref{eqscaled}, travelling wave solutions for Eq. \eqref{eqscaled2} possess  a broader transition region connecting  the two asymptotic states, $u=0$ and $u=K_1$. 

\section{Multi-kink solutions and shock-particle duality}
\label{multi kink 2}

In Section~\ref{section travelling}, we have remarked the existence of  robust localised travelling wave solutions 
for Equations \eqref{eqscaled} and \eqref{eqscaled2}. Equations    {\bf \eqref{eq:TW scaled}} and \eqref{eq:TW eq scaled 2} represent indeed kink-type  solutions which uniformly translate in both directions, according to  the sign of velocity  $c$, with respect to the chosen  reference frame in Eqs~\eqref{TraslScal} and \eqref{CRefScaling}, respectively.
 In this Section,  we prove  the existence of multi-kink solutions for Eqs \eqref{eqscaled} and \eqref{eqscaled2}, that is solutions relying on the composition of kink--like behavior.

Multi-kink solutions  describe fusion and fission of shocks, where in the process no shift in the relative phases of the colliding shocks is observed. When describing shock waves and shock-fitting procedures \cite{Whitham}, it is often useful to identify a small  parameter which permits to tune the effects of viscous and/or dispersive regularisation mechanisms. For the  Bateman-Burgers equation in conservation law form, for instance, a small parameter ahead the viscous term in the current regulates the presence of  viscosity and its impact on the solution. In the zero-viscosity limit, the  Bateman-Burgers equation turns into the Hopf equation and the behaviour of corresponding solutions  drastically changes.  When the  parameter is kept small but finite, the solutions develop instead classical (viscous) shock waves, displaying regions of high (but finite) gradient whose size can be inferred by the value of the small parameter. Smaller is the parameter, narrower is the region of  high gradient of the solution. It is therefore convenient to introduce rescalings of Eq. \eqref{GLMeq} which keep trace of the small positive parameter $\eta$, consistently with its  original derivation. Indeed, in \cite{GLM}, Eq. \eqref{GLMeq} emerged as a viscous conservation law, with  the parameter $\eta$  regulating the presence of viscosity. Detaching our analysis from any physical  derivation and interpretation of the parameter $\eta$\footnote{In \cite{GLM} and other works in the context of statistical thermodynamics, the parameter $\eta$ often arises as the inverse of the number of particles in the statistical ensemble. As a result, performing  the zero-viscosity limit o the equation turns to be equivalent to the so-called thermodynamic limit.}, we will be  dealing with two rescalings depending on the value of the quantity $\Delta=c_4 c_1^2+c_2 c_3^2 $. 

\begin{center}
	{\em Case 1: $\Delta \neq 0$.}
\end{center}
In the generic case, $c_4 c_1^2+c_2 c_3^2 \neq 0$, we  consider a set of rescaled variables 
\beq X \to \eta X \quad \text{ and } \quad T \to \eta T \,\,, \label{X T scaled eta} \eeq
where $X$ and $T$ are given by \eqref {TraslScal}, leading to the following viscous conservation law
\begin{equation}  
\partial_{T} u +     \partial_{X}             \left(\frac{1}{u} + \frac{\eta}{u}  	\partial_{T}	u	\right) = 0\,.
	\label{eqscaledv} 
\end{equation}

\begin{center}
	{\em Case 2: $\Delta= 0$.}
\end{center}
In the special case $c_4 c_1^2+c_2 c_3^2 = 0$,  we can apply a scaling transformation of the form \eqref{X T scaled eta} to the variables $X$ and $T$ are 
given by the transformation \eqref{CRefScaling}. This brings Eq. \eqref{GLMeq} into the form
\begin{equation}      \partial_{T}             \left( u + \frac{\eta}{u}  
	\partial_{X}
	u
	\right) = 0\,.
	\label{eqscaledv2} 
\end{equation}
In the following two subsections we  derive and describe multi-kink solutions to Eqs. \eqref{eqscaledv} and \eqref{eqscaledv2} separately.

\subsection{Multi-kink solutions to Equation   \eqref{eqscaledv}}

In this Section, we aim to introduce and describe multi-kink solutions to Eq.   \eqref{eqscaledv}
relying on the composition of the kinks obtained in Section \ref{section travelling}, where the number of superimposed kinks can be arbitrary. Such solutions  are obtained via the application of the   Cole-Hopf transformation  $u= \eta \partial_X \log \phi$
to an $N-$term solution of the equation \eqref{KGeq}, namely
\begin{equation}
	\label{eq:n term phi}
	\phi^{(N)}(X,T; k,\delta,\eta)= \sum_{j=1}^{N} e^{\frac{\theta_{j} (X,T; k_j, \delta_j)}{\eta}}\,\, ,
\end{equation}
where   $k=(k_1, \dots,k_N)$ and  $\delta=(\delta_1, \dots,\delta_N)$ are  $2N$ real parameters, and  
\[\theta_{j} (X,T; k_j, \delta_j ):=k_j X-\frac{ T}{k_j} + \delta_j \,\,. \] 
In order to understand the formation of multi-kink solutions, let us observe that the corresponding solution to Eq.   \eqref{eqscaled2}  can be written as follows
\begin{align}
\notag
	u(X,T;  k,\delta, \eta)&= \sum_{j=1}^{N}\frac{ k_j e^{ \theta_{j}/\eta}}{	\phi^{(N)}}\\ 
	\label{eq:weight decomposition}
	&= \sum_{ j=1}^{N}  k_j w_j (X,T; k, \delta,\eta)\,\,, 	
 \end{align}
where we have introduced the \emph{weight functions} 
\beq
w_j(X,T;  k,\delta, \eta):=\frac{e^{\theta_{j}  (X,T; k_j, \delta_j ) / \eta }}{	\phi^{(N)} (X,T; k,\delta, \eta)}, \qquad  \text{ for }\, \,  j=1,\dots,N \,\, ,
\label{eq: weights wj}
\eeq
 which satisfy the properties
\[ 0 \leq w_j(X,T; k, \delta,\eta) \leq 1\, \quad \text{and} \quad \sum_{j=1}^{N}  w_j(X,T; k, \delta,\eta) =1\,.\]

\begin{remark}
{\em 
			The decomposition of the solution  in terms of weight functions, Eq.~\eqref{eq:weight decomposition}, suggests that the potential function \eqref{eq:n term phi} can be seen as the partition function associated with a discrete probability distribution (e.g. see \cite{De Matteis biaxial} and references therein). Following this interpretation, the Cole-Hopf transformation applied to the $N-$term function \eqref{eq:n term phi} gives  the (local)  average wavenumber. Explicitly, by introducing the statistical {\em free energy} $F(X,T;k,\delta,\eta):=\eta\ln \phi^{(N)}$, we have  $u (X,T;k,\delta,\eta)\equiv   \partial_X F= \av{k}$. Higher order moments of the distribution can be inferred from higher order derivatives. For instance, the second order moment is given by $\av{k^2}=(\partial_X F)^2+ \eta \partial_X^2 F$, or, equivalently, the variance is $\text{var} (k)= \av{(k-\av{k})^2}=\eta \partial_X^2 F$.
			}
\end{remark}

Let  $\mathcal{R}_j=\{ (X,T)\in \mathbb{R}^2 \, | \, \theta_{j}  (X,T; k_j, \delta_j ) >  \theta_{i}  (X,T; k_i, \delta_i )\,, \, \forall i\neq j \}$ for $j=1, \dots, \,N$ and $\mathcal{R}=\bigcup_{j=1,\dots,N} \mathcal{R}_j$.\footnote{Observe that there might be regions $\mathcal{R}_i$ which are empty sets.
	 Therefore, for a given $N-$term solution $\phi^{(N)}$, Eq. \eqref{eq:n term phi}, the $XT-$plane is a collection of at most $N$ distinct non-empty regions $\mathcal{R}_i$.} 
The following proposition holds.

\setcounter{theorem}{0}
\begin{proposition} In the  $\eta\to 0$ limit,  solutions \eqref{eq:weight decomposition} in $\mathcal{R}$ are approximately given by
\begin{equation}
		\notag
		u(X,T;  k,\delta) \simeq k_i \,,  \quad (X,T)\in \mathcal{R}_i\,,\quad  i=1,\dots,N \,.
		\label{eq:sol geometric}
	\end{equation}\end{proposition}
\proof{Let $(X,T)\in \mathcal{R}_i$. Dividing numerator and denominator in the r.h.s. of Eq.   \eqref{eq:weight decomposition} by $e^{\theta_j/\eta}$ gives
	\begin{align}
		\notag
		u(X,T;  k,\delta)&= \sum_{ j=1}^{N} \frac{ k_j }{	1+ \sum_{l \neq j} e^{ (\theta_{l}-\theta_{j})/ \eta}} \\
		\label{eq:geometric construction}
		&\simeq k_i \,,  \quad (X,T)\in \mathcal{R}_i\,,
	\end{align}
	where we have used the fact that in $\mathcal{R}_i$ it is  $\theta_{l}-\theta_{i} <0$ for all $l\neq i$ and, as $\eta \to 0$, we have
	  $e^{(\theta_l-\theta_j)/\eta} \to \infty	 $ if $\theta_l-\theta_j>0$. 
	\footnote{Remark that the solution has an essential singularity at $\eta=0$ and therefore a standard Maclaurin expansion about $\eta$ small and positive is not possible.}
	The statement follows by varying $i=1,\dots,N$ and considering that $\mathcal{R}_i\cap \mathcal{R}_j= \emptyset$ for all  distinct pairs $\{ i, j\}$.}

Notice that, due to the linearity of functions $\theta_{j}  (X,T; k_j, \delta_j )$, regions $\mathcal{R}_j$ are bounded by line segments. Such lines emerge as resonance condition of two weight functions, that is 
\begin{equation}
		\label{eq:lines}
		\mathcal{L}_{ij}=\{ (X,T)\in \mathbb{R}^2 \, | \, \theta_{i}  (X,T; k_i, \delta_i ) =\theta_{j}  (X,T; k_j, \delta_j )>  \theta_{l}  (X,T; k_l, \delta_l )\,, \, \forall l\in \{1, \dots,N \}\setminus\{i,j\}   \}\,,
\end{equation}
and represent the trajectories of classical, viscous, shock waves. In particular, the slope associated with the line $\mathcal{L}_{ij}$ is given by $v_{ij}=- (k_i k_j)^{-1}$ and provides the velocity of propagation of the corresponding shock. 
Resonances between $n>2$ weight functions are in principle allowed and occur at specific points in the $XT$-plane.  
 More specifically, the  location of the resonance  between  three adjacent asymptotic states  $k_i$, $k_j$ and $k_l$ is  defined by the system of linear equations $\theta_i(X,T;k_i,\delta_i)=\theta_j(X,T;k_j,\delta_j)=\theta_l(X,T;k_l,\delta_l)$. 
 Notice  that this system   admits a unique  solution, given that all $k_j$ are distinct and this is so for the shock velocities. Explicitly, by introducing 
$\overline{\kappa}_n = \lf k_i^n, k_j^n, k_l^n\rg^\mathbf{\top}$ and $\overline{\delta} = \lf \delta_i, \delta_j, \delta_l\rg ^\mathbf{\top}$, the three-term resonance occurs at\footnote{Explicitly, we have 
\[ X_{ijl} =\frac{ k_i k_j
				(\delta_j-\delta_i)+k_i k_l ( \delta_i-\delta_l)+k_j k_l (\delta_l-\delta_j)}{( k_i- k_j) (k_i-k_l) (k_j-k_l)} \,\, , \qquad 
				T_{ijl}= \frac{ k_i k_j k_l \left(k_i ( \delta_l-\delta_j)+k_j (\delta_i-\delta_l)+k_l (\delta_j-\delta_i)\right)}{(k_i-k_j) (k_i-k_l) (k_j-k_l)} \,\, \,.
\]
}
	\begin{equation} X_{ijl}= \frac{ \det \lfq ( \overline{\kappa}_{0},  \overline{\kappa}_{-1}, \overline{\delta})\rgq}{ \det \lfq  (\overline{\kappa}_{0},  \overline{\kappa}_{1},\overline{\kappa}_{2} )\rgq} \, k_ik_jk_l \, \, , \qquad
		T_{ijl}= \frac{ \det \lfq  (\overline{\kappa}_{0},  \overline{\kappa}_{1}, \overline{\delta})\rgq}{ \det \lfq  (\overline{\kappa}_{0},  \overline{\kappa}_{1},\overline{\kappa}_{2} )\rgq} \,  
		\, k_ik_jk_l \,\, ,
		\label{vertex}
	\end{equation}
	where the \emph{Vandermonde determinant}  appearing  at the denominator is nonzero. 

One can also  look for higher order resonances, which will eventually lead to points where three or more lines intersect. In general, if the intersection occurs at resonances of $n$ weight functions $w_{i}\,, \dots, w_{j}$, then solution \eqref{eq:weight decomposition} at the intersection point  can be approximated by
\[ 	u(X,T;  k,\delta) \simeq \frac{k_i + \dots + k_j}{n} \,,
\]
that is the value of  $u(X,T)$ is approximately given by  arithmetic mean of the corresponding asymptotic states. 
It is also worth to stress that the for a given $N-$term function $\phi^{(N)}$, the $XT-$plane is a collection of at most $N$ distinct regions $\mathcal{R}_j$. Formula \eqref{eq:sol geometric} can be seen as the asymptotic solution in the regime of small viscosity $\eta$ and is actually a time-dependent  generalised \emph{Heaviside step function}.

\setcounter{theorem}{5}
\begin{remark}  
	{\em The scheme adopted above to characterise the multi-kink behaviour has been to design a {\em skeleton}  in the space of independent variables corresponding to different asymptotic values of the dependent variable. The resulting approximation in the $\eta \to 0$ limit, Eq.   \eqref{eq:geometric construction},  could be also equivalently seen as  the {\em tropical limit}   of the $N-$term solution \eqref{eq:weight decomposition} (see e.g. \cite{kato,Athorne}  and Refs. therein).
Indeed, more standardly, one may perform the $\eta \to 0$ approximation  at the level of the potential function \eqref{eq:n term phi}, resulting into
		 $\phi^{(N)}(X,T;k,\delta,\eta) \simeq \max_{j}  \,\{ e^{\theta _j/ \eta} \}\,, \quad (X,T)\in \mathcal{R}_j.$
		 The application of the Cole-Hopf transformation to the approximated potential function would still give $u(X,T;k,\delta,\eta)\simeq k_j$ in $\mathcal{R}_j$.
		}
		
\end{remark}

The decomposition of the solution \eqref{eq:n term phi} in terms of elementary single-kink solutions, can be understood as follows. Let $\mathcal{R}_i$ and $\mathcal{R}_j$ be two adjacent regions characterised by parameters $k_i$, $\delta_i$ and $k_j$, $\delta_j$, respectively. In the region $\mathcal{R}_{ij}=\mathcal{R}_i \cup \mathcal{R}_j\cup \mathcal{L}_{ij}$, we have that 
	\begin{align}
		\notag
		u(X,T;k,\delta) &=  \frac{ k_i e^{\theta_{i}/ \eta} +k_j e^{\theta_{j}/ \eta} }{	\left( e^{\theta_{i}/ \eta}+e^{\theta_{j}/ \eta} \right) \left(1+\sum_{l\neq i,j } \left(e^{(\theta_i-\theta_l)/\eta} + e^{(\theta_j-\theta_l)/\eta} \right)^{-1} \right)} 
		\\   \notag &  \qquad \qquad \qquad \quad
		+\sum_{s\neq i,j}^{N} \frac{ k_s \left( e^{ (\theta_{s}-\theta_{i})/ \eta}+ e^{ (\theta_{s}-\theta_{j})/ \eta} \right)}{1+\sum_{l\neq i,j } \left(e^{(\theta_i-\theta_l)/\eta} + e^{(\theta_j-\theta_l)/\eta} \right)^{-1}}\\
		\label{eq:multi-kink single-kink approx}
		& \qquad  \qquad \qquad \quad\simeq  \frac{ k_i e^{\theta_{i}/ \eta} +k_j e^{\theta_{j}/ \eta} }{	e^{\theta_{i}/ \eta}+e^{\theta_{j}/ \eta}  }= \frac{k_i+k_j}{2} +  \frac{k_i-k_j}{2} 
		\tanh{\left( \frac{\theta_i -\theta_j}{2\eta}\right)} \,\, ,
	\end{align} 
	where we have used the fact that for $(X, T)\in \mathcal{R} _{ij}$ it is $\theta_{l}-\theta_{i}<0$ and $\theta_{l}-\theta_{j}<0$  for all $l\neq i,j$, and as $\eta \to 0$ we have $e^{(\theta_{l}-\theta_{i})/\eta}\to 0$ and $e^{(\theta_{l}-\theta_{j})/\eta} \to 0$.
	The approximation in Eq. \eqref{eq:multi-kink single-kink approx} is just the single-kink solution found in Eq. \eqref{eq:TW scaled}, upon identifying the kink speed with $c=-(k_i k_j)^{-1}$ and  integration constants as 
	\begin{equation} K_1 =\sqrt{\left( \frac{k_i}{2} \right)^2+ \left( \frac{k_j}{2}\right)^2 } \quad , \quad K_2= \frac{\delta_i-\delta_j}{k_i-k_j}\,. 
	\end{equation}

 	\begin{remark}  {\em
	Formula \eqref{eq:multi-kink single-kink approx} states that, far from the kink-kink interaction, the properties of the kink \eqref{eq:multi-kink single-kink approx} in $ \mathcal{R} _{ij}$ are fully characterised by the  pair  $\{e^{\theta_i /\eta}, e^{\theta_j/\eta}\}$, hence implying that no phase--shift results from the interaction with the other kinks.
	}
	\end{remark}

In the next subsection we wish to describe the shock dynamics displayed by the multi-kink solutions \eqref{eq:weight decomposition} under the light of multiple scattering processes among  particles and associated conservation laws. This will allow us to establish a shock-particle duality for the underlying model \eqref{eqscaledv}.

\subsubsection{Shock--particle duality} 
Before analysing the shock dynamics, we should first identify the \emph{mass} and \emph{momentum} associated with the conservation law. Let $-\partial_T u (X,T)$ be the \emph{momentum density} at time $T$ and location $X$. The total momentum at a given time $T$ is given by
\begin{equation}
	\label{eq:momentum global}
	P(T):= -\int_{-\infty}^\infty \,\partial_T u \, dX =   \int_{-\infty}^\infty \, \p_X \left(  \frac{1}{u} +\frac{\eta}{u}  \partial_T u \right)\, dX =  \frac{1}{u} +\frac{\eta}{u}  \partial_T u \, \Big|_{-\infty}^{+\infty}  = \frac{k_{-}-k_{+}}{k_{-}\, k_{+}}\,\, ,
\end{equation}
where $k_-$ and $k_+$,  with $k_{\pm}\in\{ k_1, \dots, k_N\}$, are precisely the asymptotic values of the solution at that time $T$, that is 
\[k_{\pm}:=\lim_{X\to \pm \infty} u(X,T)\,\,.\]
In Eq. \eqref{eq:momentum global}, we have used the fact that $\partial_T u \lf X, T \rg \to 0$ for $X \to \pm \infty$ at any time $T$. 
Let $\partial_X u$ be the \emph{mass density}. The total \emph{mass} at time $T$ is given by
\begin{equation}
	\notag
	M(T):=\int_{-\infty}^{+ \infty} \partial_X u \,  dX =  k_{+}-k_{-}\,\,.
\end{equation}
Notice that  total \emph{mass}, \emph{momentum} and \emph{velocity} are consistently related through the relation $P=M v_{\pm}$, where $v_{\pm}=-(k_+ k_-)^{-1}$ is the velocity of a shock between the most outer states, $k_{\pm}$.
Remarkably, mass and momentum are conserved quantities for this system, that is $P'(T)=M'(T)=0$ (the prime meaning now the $T$-derivative), and conservations of mass and  momentum emerge as for a  scattering process. More explicitly, we have 
\begin{align}
	\label{eq:conserv of mass}
	\sum_{\{i,j\}} m_{ij}^{(\text{in})}&= \sum_{\{i,j\}} m_{ij}^{(\text{out})} \\
	\label{eq:conserv of momentum}
	\sum_{\{i,j\}} m_{ij}^{(\text{in})}v_{ij}^{(\text{in})}&= \sum_{\{i,j\}} m_{ij}^{(\text{out})}v_{ij}^{(\text{out})}\,,
\end{align}
where `in' and `out' label mass and velocities of the shocks  before and after scattering\footnote{A subtle remark is needed here. Actually, one retrieves scattering among classical (point-like) particles in the limit $\eta \to 0$, that is when mass and momentum are strongly localised about the centre of each shock. For $\eta$ small but finite, and outside the interaction region, one can simply identify the mass and momentum carried by each shock by performing an integral over a domain of size of order $O(\eta)$ centred about the shock location. }, respectively, and the summation is performed over all pairs of   wavenumbers involved in the scattering. In Eqs. \eqref{eq:conserv of mass}-\eqref{eq:conserv of momentum}, we have identified the mass
 and the momentum carried by each individual shock  as $m_{ij}=k_j -k_i$  and $p_{ij}=m_{ij} v_{ij}$. 
Conservation of mass is nonetheless  the analogous of Rankine-Hugoniot jump condition  for a fluid supporting discontinuous shock waves \cite{Whitham}. From this point of view, the equation of interest \eqref{GLMeq}, or  \eqref{eqscaledv}, represents viscous regularisation \cite{arsie} of the continuity equation by the flow density  
\[ Q\lf u \rg \to Q\lf u \rg  + \eta R\lf u \rg \partial_T u\,,\]
where  here $Q\lf u \rg =  R\lf u \rg = 1/u $.
This procedure is different from the usual Burgers regularisation, $Q\lf u \rg \to Q\lf u \rg  +   \eta  R\lf u \rg \partial_X u$, when $Q\lf u \rg = u^2$ and $R(u)=1$, or dispersive  ones  
(see e.g. \cite{El} and references therein). We will discuss in detail  the shock dynamics, including the conservation laws \eqref{eq:conserv of mass} and  \eqref{eq:conserv of momentum}, with the aid of some examples. 
\begin{figure}[h!]
	\centering 
	\includegraphics[scale=0.5,angle=0]{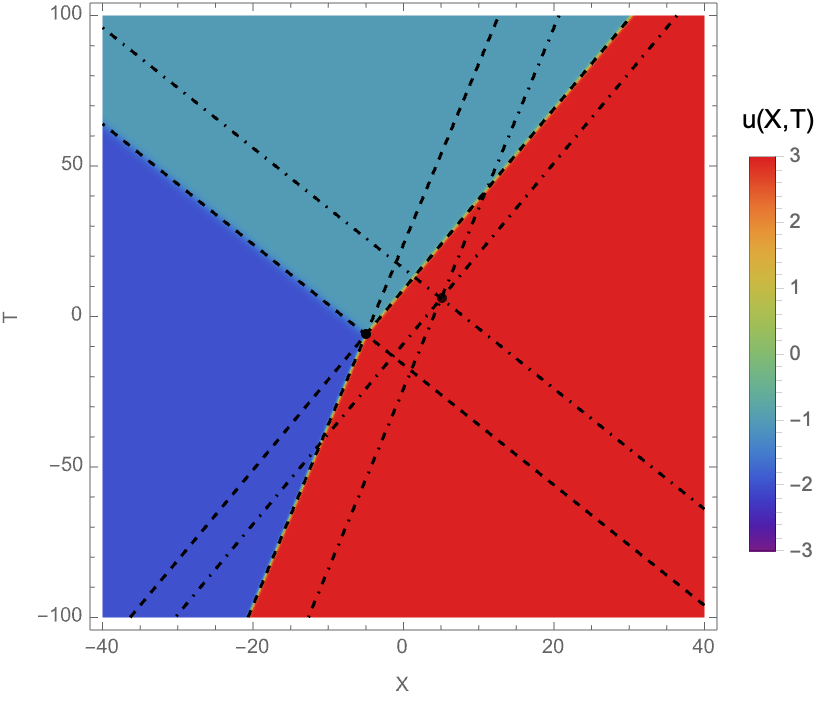}  
	\includegraphics[scale=0.5,angle=0]{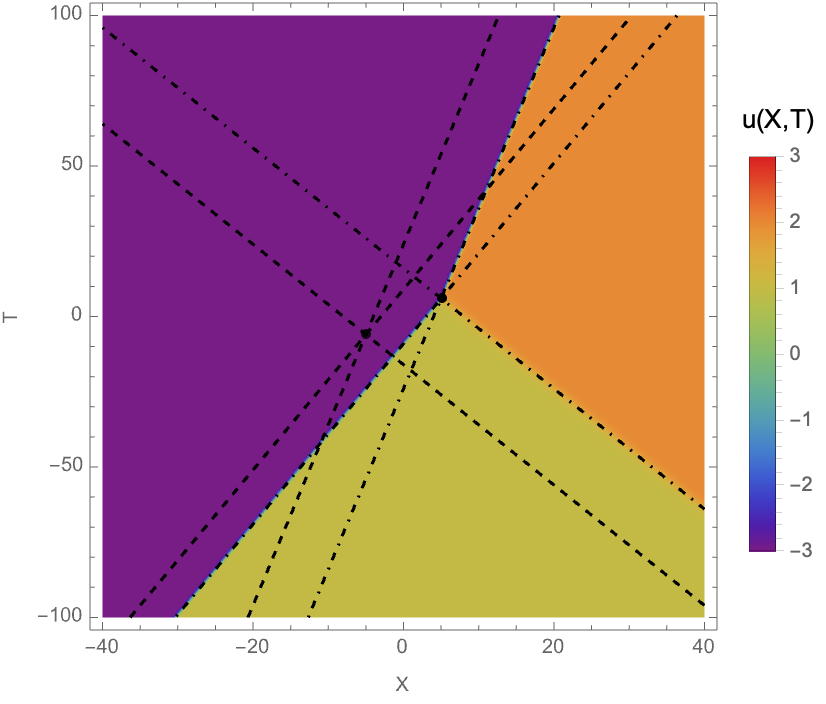}  
	\caption{\small{  A pair of $\widetilde{\mathcal{PT}}$--connected solutions representing  fission and fusion of two shocks. 	Parameters are set to $k_1=-2, \, k_2=-1,\, k_3=3$,  
					$\delta_1=0,\, \delta_2=8,\, \delta_3=20$ and $\eta=1/2$.   \emph{(Left)} The fission process of a shock wave    decaying in two shocks.  \emph{(Right)} The $\widetilde{\mathcal{PT}}$-- transformed solution representing two shocks   merging into a single shock.  The dashed and dash--dotted semi-lines of resonance are  obtained in both processes as conditions $\theta_i(X,T; k_i,\delta_i)=\theta_j(X,T; k_j,\delta_j)$ for all pairs of distinct indices $i,j=1,2,3$. }
				\label{fig: fission-fusion}	} 
\end{figure} 

We first start with the case of $n=3$ terms in Eq.~\eqref{eq:n term phi}. This case represents indeed the elementary fusion or fission of shocks, as displayed in Figure~\ref{fig: fission-fusion} where a relatively large value of $\eta$ has been chosen to magnify the finite-size region of high gradient.  The Left panel shows  the fission process of a shock wave entering at $T\to -\infty$ with velocity $v_{13}= 1/6 $ and mass $m_{13}=5$ decaying in two shocks of velocities and masses  equal to  $v_{12}=-1/2$, $v_{23}=1/3$  and $m_{12}=1$,  $m_{23}=4$, respectively.  The Right panel exhibits instead  the $\widetilde{\mathcal{PT}}$-- transformed solution representing two shocks of velocities and masses equal to $v_{32}=1/3$, $v_{21}=-1/2$ and $m_{32}=4$, $m_{21}=1$ merging into a single shock with mass and velocity $m_{31}=5$ and $v_{31}=1/6$, respectively. 
Notice that asymptotic states in the Right panel take the values $\widetilde{\mathcal{PT}} k_1=-k_1=2$, $\widetilde{\mathcal{PT}} k_2=-k_2=1$ and $\widetilde{\mathcal{PT}} k_3=-k_3=-3$. The location and time of collision for the processes  (highlighted with  black circles) are computed by formula \eqref{vertex} giving $(X_{123},T_{123})=(-5,-6)$ and  $\widetilde{\mathcal{PT}}(X_{123},T_{123})=(5,6)$, respectively.  
Semi-lines of resonance ${\cal L}_{1 3}$, ${\cal L}_{2 3}$ and ${\cal L}_{1 2}$,  defined by Eq.~\eqref{eq:lines}, emerge from the dashed and dash--dotted lines  simply obtained in both processes as conditions $\theta_i(X,T; k_i,\delta_i)=\theta_j(X,T; k_j,\delta_j)$ for all pairs of indices $i,j=1,2,3$.

Fusion and fission processes are reversible due to the $\mathcal{PT}$--like symmetry enjoyed by Eq. \eqref{eqscaledv}. For instance, the  fusion process of two shocks in the right panel  is obtained by application of  the  transformation $\widetilde{\mathcal{PT}}$ defined in Eq. \eqref{eq:PT symmetry} to the solution representing the fission of two shocks displayed in the left panel. The resulting processes resemble those  of the  fission of a particle of mass $m_{13}=5$ into two particles of masses $m_{12}=1$ and $m_{23}=4$, and the fusion of two particles of masses $m_{21}=1$ and $m_{32}=4$ to give a particle of mass $m_{31}=5$, respectively.

\begin{figure}[h!]
	\centering 
	\includegraphics[scale=0.5,angle=0]{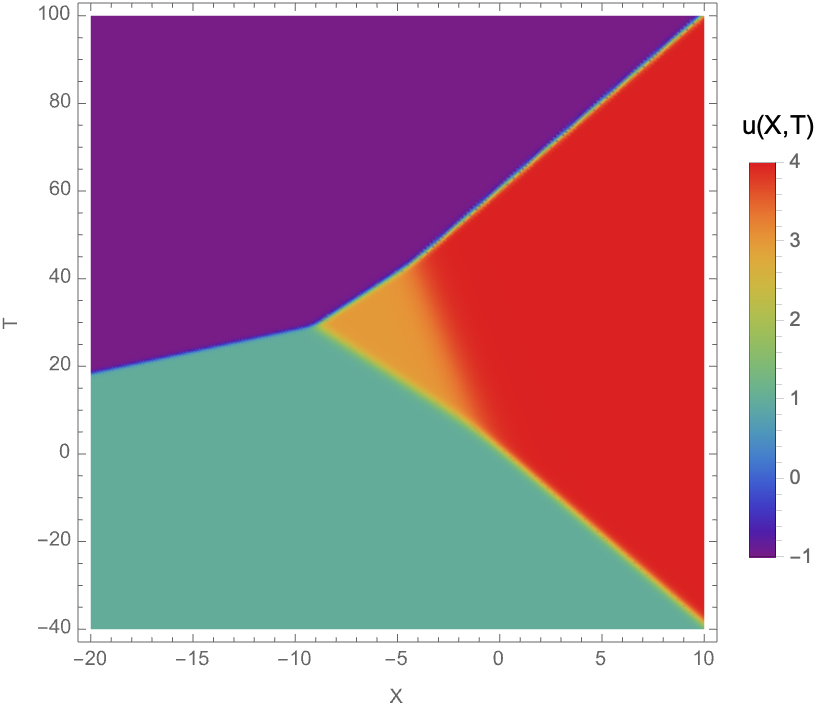}  
	\includegraphics[scale=0.5,angle=0]{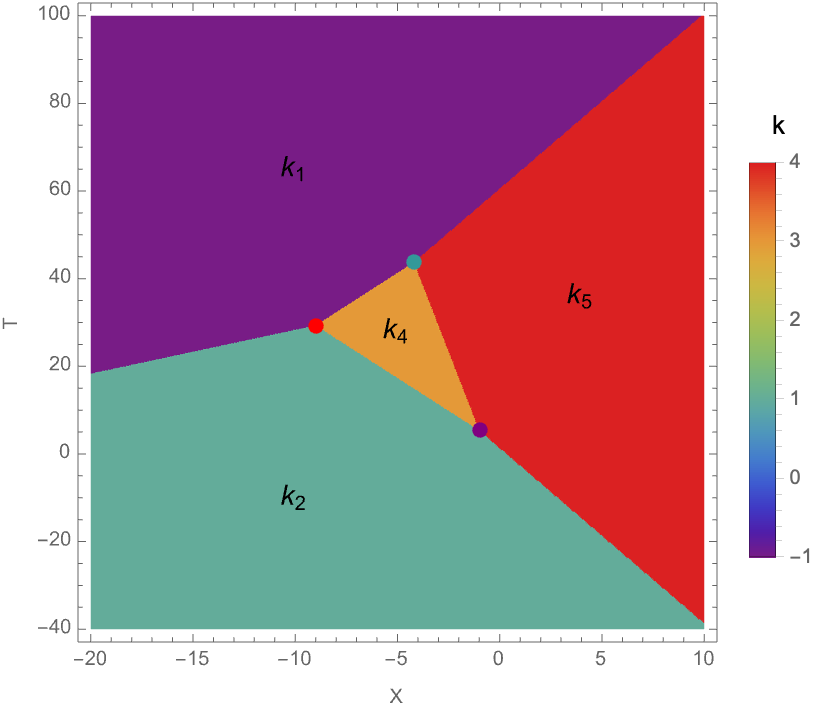}  
	\caption{\small{Multi-kink solution~\eqref{eq:weight decomposition} with $N=5$. Parameters are set to $k_1=-1, \, k_2=1,\, k_3=2,\, k_4=3\,, k_5=4$,  $\delta_1=-227/3,\, \delta_2=1,\, \delta_3=-20/3,\,\delta_4=-5/9\,, \delta_5=0$ and $\eta=0.5$.  \emph{(Left)} Solution for $u(X,T)$.   \emph{(Right)} Geometric construction identifying the asymptotic, discontinuous solution \eqref{eq:geometric construction}.} } 
	\label{fig: 5coll}
\end{figure}

Let us now analyse  Figure \ref{fig: 5coll}, which was obtained from $n=5$ terms in Eq.~\eqref{eq:n term phi}. On the right, our geometric construction, which we refer as asymptotic solution, displays four different regions $\mathcal{R}_1$,  $\mathcal{R}_2$, $\mathcal{R}_4$ and $\mathcal{R}_5$ characterised by wavenumber $k_1=-1$, $k_2=1$, $k_4=3$ and  $k_5=4$ separated by six   line segments: $\mathcal{L}_{12}$, $\mathcal{L}_{25}$,  $\mathcal{L}_{24}$, $\mathcal{L}_{45}$, $\mathcal{L}_{14}$  and  $\mathcal{L}_{15}$. The corresponding multi-kink solution for $u(X,T)$ is displayed to the left, showing agreement with the  geometric construction. Each line segment  correspond to the trajectory of a kink, thought as a classical viscous shock wave. Shock fission, i.e. generation of multiple kinks from a single one, occurs when two lines originate from one.  Shock confluence or fusion occurs instead when two shocks collide to form a single shock.  The dynamics of shocks can be understood in terms of  \emph{inelastic scattering}, that is particle-particle collisions where new particles are formed from old ones while preserving the overall mass and momentum.  Shock fission can be understood as a particle decay.
 Similarly, confluence of  shocks is understood as fusion of  particles. 
Conservation of mass and momentum are therefore expressed in terms of conservation of mass and momentum of shocks. Direct application of conservation laws \eqref{eq:conserv of mass}-\eqref{eq:conserv of momentum} to the case of a fission of one shock to form a pair of shocks gives
\begin{align}
	\label{eq:conservation of mass}
	k_i-k_j&= k_i-k_l+ k_l-k_j \,\, , \\
	\label{eq:conservation of momentum}
	\frac{k_i-k_j}{k_i k_j}&=\frac{k_i-k_l}{k_i k_l}+\frac{k_l-k_j}{k_l k_j}\,\,. 
\end{align}
Eqs.~\eqref{eq:conservation of mass} and \eqref{eq:conservation of momentum} can be straightforwardly generalised to other shock fusion/fission processes.  Relatively to the example under study, the shock between the regions $\mathcal{R}_2$ and  $\mathcal{R}_5$, which  is approximately   $m_{14}=3$ in amplitude (or \emph{mass}), performs a fission to two shocks: one, to the left, having amplitude  $m_{24}=2$ and another one,  to the right, having amplitude $m_{45}=1$. Therefore the mass of the incoming shock is distributed across the two outgoing shocks such that mass is conserved in the scattering process. The same happens for   momentum. In fact, from \eqref{eq:momentum global} one deduces 
$P=-5/4$ for any  value of $T$, given that the asymptotic states at $X\to \pm \infty$ are the same at all times.

Asymptotically, the overall process can be described as a fission of two shocks with three intermediate shocks enclosing a transient state ($u(X,T)\simeq k_4=3$). In this example, shocks interact forming  a network of $Y-$shaped trajectories \cite{Kodamabook}. Two fissions and one fusion are displayed at $(X,T)=(X_{245},T_{245}), \,(X_{124},T_{124})$ and $(X_{145},T_{145})$, respectively. 
\begin{figure}[h!]
\centering 
\includegraphics[scale=0.5,angle=0]{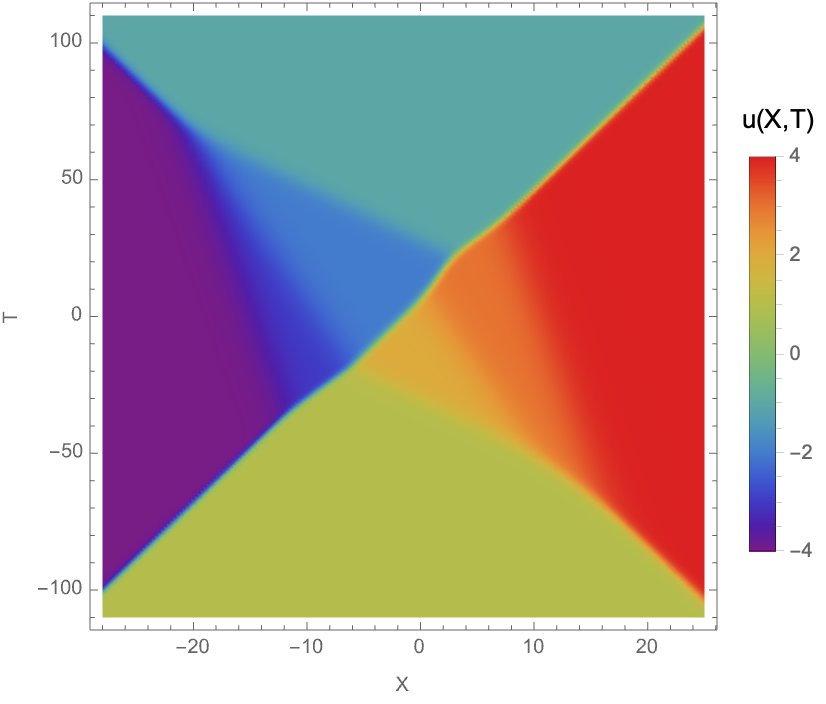}  
\includegraphics[scale=0.5,angle=0]{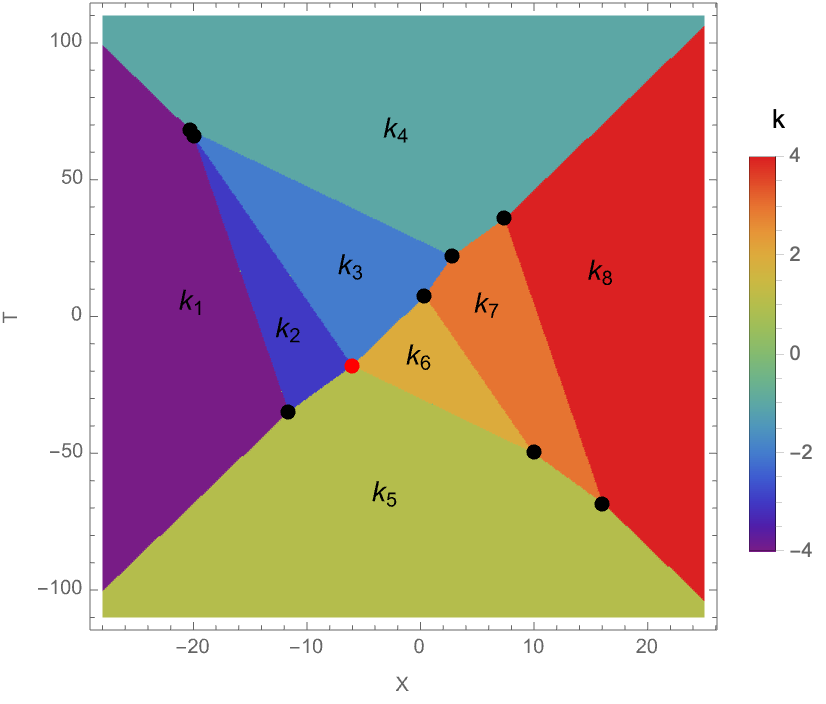}  
\caption{\small{Multi-kink solution~\eqref{eq:weight decomposition} with $N=8$. Parameters are set to $k_1=-4, \, k_2=-3,\, k_3=-2,\, k_4=-1 \,, k_5=1\,,k_6=2\,,k_7=3\,,k_8=4$,  
		$\delta_1=-29/2,\, \delta_2=0,\, \delta_3=9,\,\delta_4=-115/24\,, \delta_5=0,\,\delta_6=15,\,\delta_7= 107/8,\,\delta_8=3$ and $\eta=1$.  \emph{(Left)} Solution for $u(X,T)$.   \emph{(Right)} Geometric construction identifying the asymptotic, discontinuous solution, $u(X,T)\simeq k_j$, for $(X,T)\in \mathcal{R}_j$. Multiple fusion and fission of shocks are displayed, including the simultaneous fusion and fission (red circle) of two initial shocks to give two new ones.  }} 
\label{fig: 8coll}
\end{figure} 

\begin{figure} 
\centering 
\includegraphics[scale=0.71]{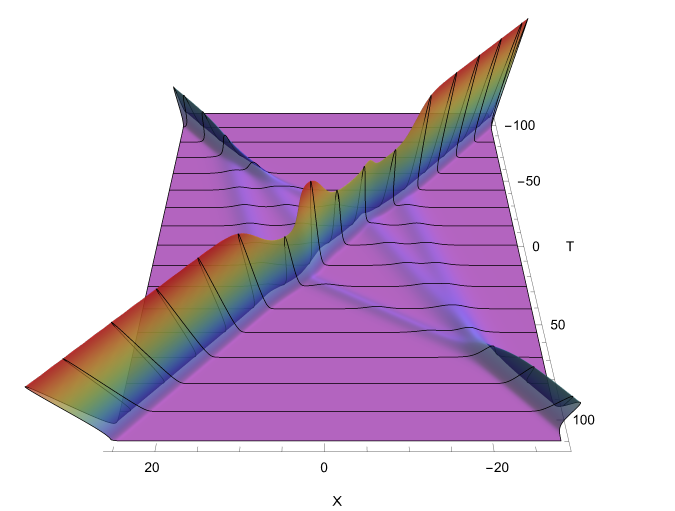}
\includegraphics[scale=0.48,angle=0]{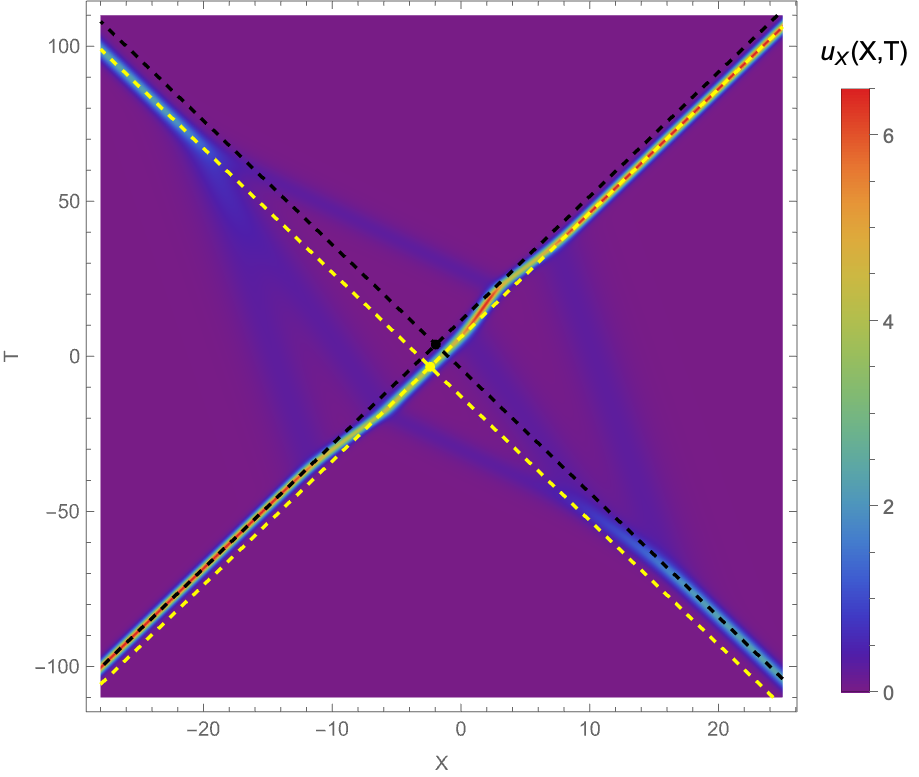} 
\caption{\small{Mass density $u_{X}(X,T)$ for the multi-kink solution~\eqref{eq:weight decomposition} with $N=8$. Parameters are the same as in Fig~\eqref{fig: 8coll},  $k_1=-4, \, k_2=-3,\, k_3=-2,\, k_4=-1 \,, k_5=1\,,k_6=2\,,k_7=3\,,k_8=4$,  
		$\delta_1=-29/2,\, \delta_2=0,\, \delta_3=9,\,\delta_4=-115/24\,, \delta_5=0,\,\delta_6=15,\,\delta_7= 107/8,\,\delta_8=3$ and $\eta=1$.  Both figures show the mass density $u_{X}(X,T)$ and its soliton behaviour. Multiple fission and fusion of solitons are displayed in the \emph{Left} panel.  The density plot on the \emph{Right} aims instead  at showing the effective phase shift due to the interaction. The black dashed lines   are constructed along   the centres of  the trajectories of   the incoming shocks, while the  yellow dashed lines are instead obtained from the  trajectories  of the outgoing shocks. The black and yellow circles highlight the intersections between ingoing and outgoing lines.   }} 
\label{fig: 8coll grad}
\end{figure}

A more complex scenario emerges in  Fig~\eqref{fig: 8coll}, where an example of a  $N=8-$term solution is considered.
  The overall process results in a scattering between two initial shocks of mass and momentum $m_{15}=5$, $m_{58}=3$ and  $p_{15}=5/4$, $p_{58}=-3/4$, respectively, to give two final  shocks  of mass and momentum $m_{14}=3$, $m_{48}=5$ and  $p_{14}=-3/4$, $p_{48}=5/4$. However, four metastable states, labelled with $k_7$, $k_6$,  $k_2$ and $k_3$, appear. Remarkably, the two shocks between states $k_2$ and $k_5$, and $k_5$ and $k_6$ collide at $(X_{2356},T_{2356})=(-6,-18)$ to form two new shocks between states $k_2$ and $k_3$, and $k_6$ and $k_3$ (red circle in Fig~\eqref{fig: 8coll}). Notice that mass and momentum are conserved through the collision. It is also important to remark that the asymptotic states, that is the solution at  large positive and negative times, could be described by two colliding particles, one moving from  left to right  with mass $m_{15}\equiv m_{48}=5$ and velocity $v_{15}\equiv v_{48}=1/4$, and another one with  mass $m_{58} \equiv m_{14}=3$ and velocity $v_{58} \equiv v_{14}=-1/4$. The effect of the collision results in an effective phase-shift, that is the two initial particles are either delayed  or advanced compared to the case of non-interacting particles. Indeed, extending the trajectories of the four asymptotic  shocks to the central, transient region, one notices that the trajectories do not intersect in a single point. This situation can be better analysed by looking at the behaviour of $u_X(X,T)$, as displayed in Fig~\eqref{fig: 8coll grad}.  The two set of  lines are indeed parallel in pairs, meaning that the velocities of the ingoing and outgoing shocks are the same. However,  
the fact that lines do not coincide suggests the presence of an effective phase shift. In fact, the intersection points, thought as the centres of the scattering in the linear interaction regime, are given by  $(X^{\text{(in)}},T^{\text{(in)}})=(-39/20,19/5)$ and $(X^{\text{(out)}},T^{\text{(out)}})=(-863/360,-151/45)$, respectively. The phase-shift of both shock waves produces a shift in position, $\left( \Delta_{\rightarrow} ,  \Delta_{\leftarrow} \right) \simeq (0.7,-1.6)$, which is  given by a Galilean transformation
with  $\Delta_{\rightarrow/\leftarrow}=X^{(\text{out})}- X^{(\text{in})} -\left( T^{(\text{out})}- T^{(\text{in})}  \right)  v_{\rightarrow/\leftarrow} $, where subscripts `$\rightarrow$' and `$\leftarrow$' label the shocks propagating left-to-right and right-to-left, respectively.  Away from the interaction, shocks travel with uniform  rectilinear motion as they were free particles. Indeed, one can notice that the interaction between the incoming shocks start to occur at about $T=-60$ and seems to be negligible for $T>60$. For $-60\lesssim  T\lesssim 60$, that is  for $-20\lesssim X \lesssim 20$, an interesting dynamics emerge, where multiple intermediate shocks are formed, traced by peaks in the gradient. Notice that higher is the peak, larger is the  mass transported by the shock. 

\subsection{Multikink solution for Equation \eqref{eqscaledv2}}
The arguments leading to the derivation and analysis of multi-kink solutions of Eq.~\eqref{eqscaledv2} slightly differ from those concerning Eq.~\eqref{eqscaledv}. Indeed,  the general solution \eqref{sol Delta 0}  poses a severe limitation on the structures that can be employed while conceiving solutions displaying kinks. In practice, 
\eqref{sol Delta 0} and its counterpart with rescaled variables, Eq.~\eqref{eqscaledv2}, is  compatible with  the presence of a nonlinear superposition of  terms in the form  $e^{(k_jX -\omega(k_j) T +\delta_j )/\eta }$, where, differently from Eq.~\eqref{eqscaledv}, terms with the null-momentum are allowed and obey to a distinct dispersion relation. Without loss of generality, let $k_0=0$ and $k_j\neq 0$ for $j>0$. Multi-kink solutions to Eq.~\eqref{eqscaledv2} take the form 
\beq
u(X,T)=\eta \, \partial_X \ln \left(\sum_{j=0}^N e^{\theta^{(0)}_j/\eta} \right )  \,\, , \qquad \theta^{(0)}_j:=k_j X- \omega_j T+\delta_j  \,\, , 
\label{eq N kink Delta0}
\eeq  
with $\omega_0=0$ and $\omega_j=\omega^{(0)}\in\mathbb{ R}$ for $j>0$. Similarly to the case $\Delta\neq 0$, we can identify domains $\mathcal{R}_j^{(0)}$ in which $\theta_{j}^{(0)}$ are dominant, that is  $\mathcal{R}_j^{(0)}=\{ (X,T)\in \mathbb{R}^2 \, | \, \theta_{j}^{(0)}  (X,T; k_j, \delta_j ) >  \theta_{i}^{(0)} (X,T; k_i, \delta_i )\,, \, \forall \, i\neq j \}$ for $j=0, \dots, \,N$ and $\mathcal{R}^{(0)}=\bigcup_{j=0,\dots,N}\mathcal{R}_j^{(0)}$. Regions $\mathcal{R}_j^{(0)}$ are bounded by line segments which are now of two types depending on the wavenumbers characterising the two adjacent states. More precisely, the line segments 
\[\mathcal{L}_{ij}^{(0)}=\{ (X,T)\in \mathbb{R}^2 \, | \, \theta_{i}^{(0)}  (X,T; k_i, \delta_i ) =\theta_{j}^{(0)}  (X,T; k_j, \delta_j )>  \theta_{l} ^{(0)} (X,T; k_l, \delta_l )\,, \, \forall \, l\in \{0, \dots,N \}\setminus\{i,j\}   \}\,\] 
identify trajectories of viscous shocks, which are either stationary, that is with velocity $v_{ij}=0$, if  $i,j>0$, or move with velocity $v_{i0}=\omega^{(0)}/k_i$ if  $i>0$. 
Similarly to solutions to Eq.~\eqref{eqscaledv}, the following proposition is implied.

\setcounter{theorem}{1}
\begin{proposition} In the limit $\eta\to 0$, solutions \eqref{eq N kink Delta0} in $\mathcal{R}^{(0)}$   are approximately given by
\begin{equation}
	\notag
	u(X,T;  k,\delta) \simeq k_j \,,  \quad (X,T)\in \mathcal{R}_j^{(0)}\,,\quad  j=0,\dots,N \,.
	\label{eq:sol geometric d0}
\end{equation}
\end{proposition}
\proof{The proof is obtained by introducing $N+1$ \emph{weight functions} and is identical to that in Proposition 1. }

\begin{figure}[h!]
\centering 
\includegraphics[scale=0.49,angle=0]{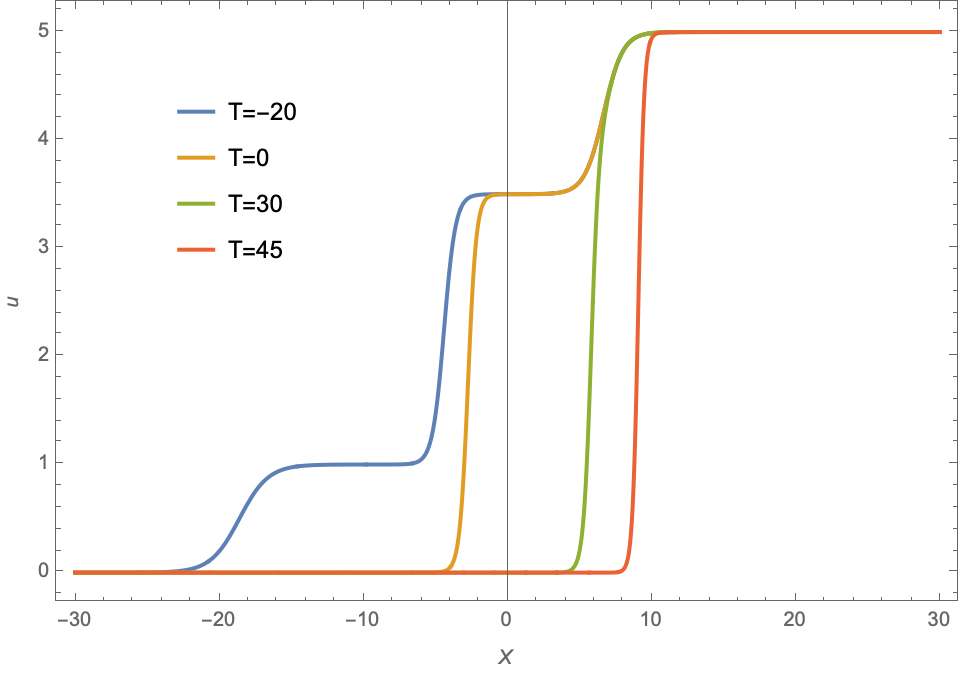} \,\, 
\includegraphics[scale=0.475,angle=0]{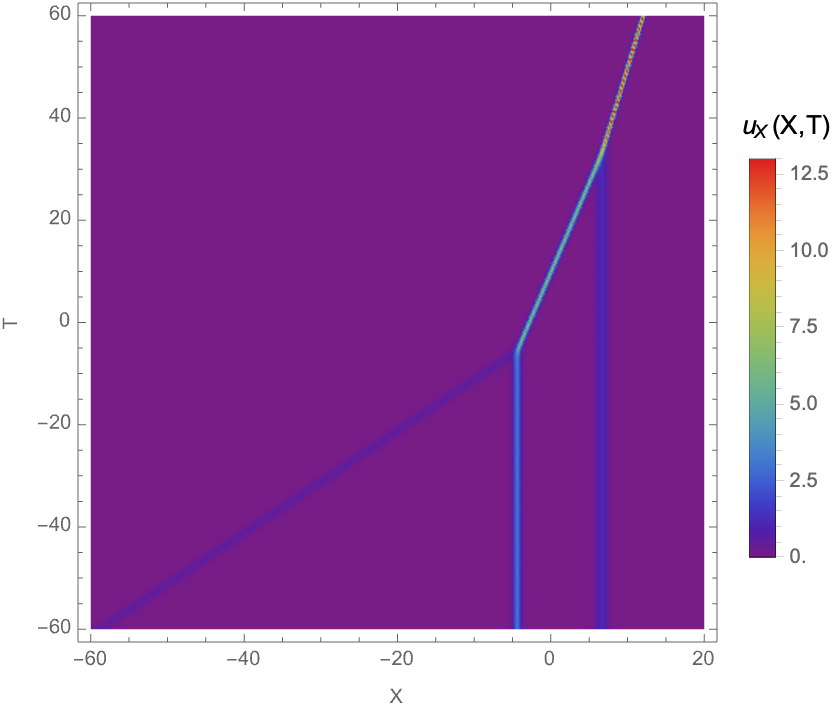}
\caption{\small{
	Solution~\eqref{eq N kink Delta0} with four terms,  $k_0=0$, $k_1=3.5, \, k_2=1,\, k_3=5,\, \delta_0=\ln{3/2}, \delta_1=10 \,, \delta_2=-1\,,\delta_3=0\,,A_0=1.5$, $\omega_0=1$ and $\eta=1$.
	\emph{(Left)} Solution at fixed times $T=-20,0,30, 45$ evidencing a fusion mechanism. 
	\emph{(Right)} Gradient of the solution, $u_X(X,T)$, evidencing the shock trajectories and fusion  under the light of scattering processes.} 
	}
	\label{fig: N3 Delta0}
\end{figure}

\begin{figure}[h!]
\centering 
\includegraphics[scale=0.5,angle=0]{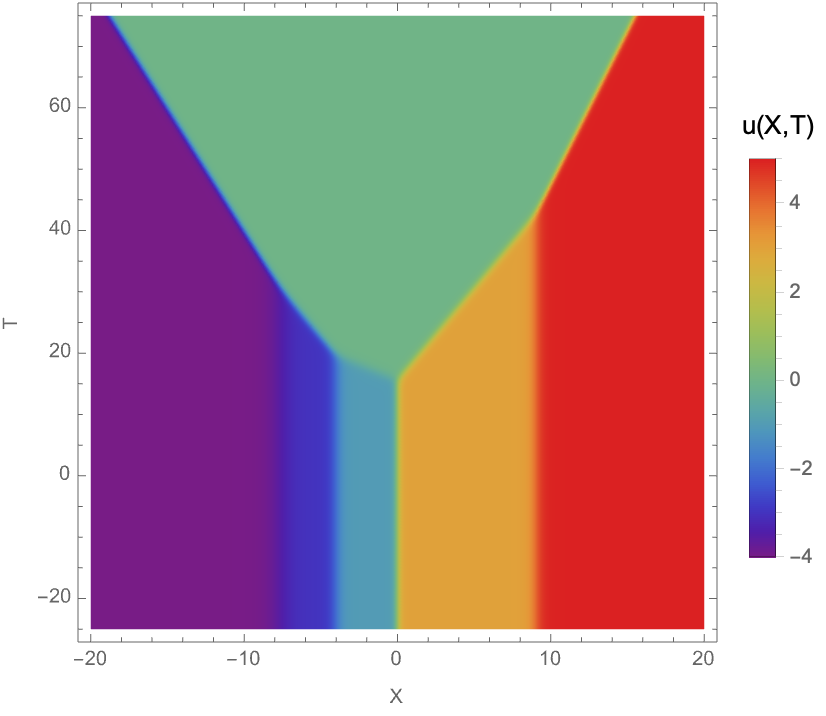} \,\, 
\includegraphics[scale=0.5,angle=0]{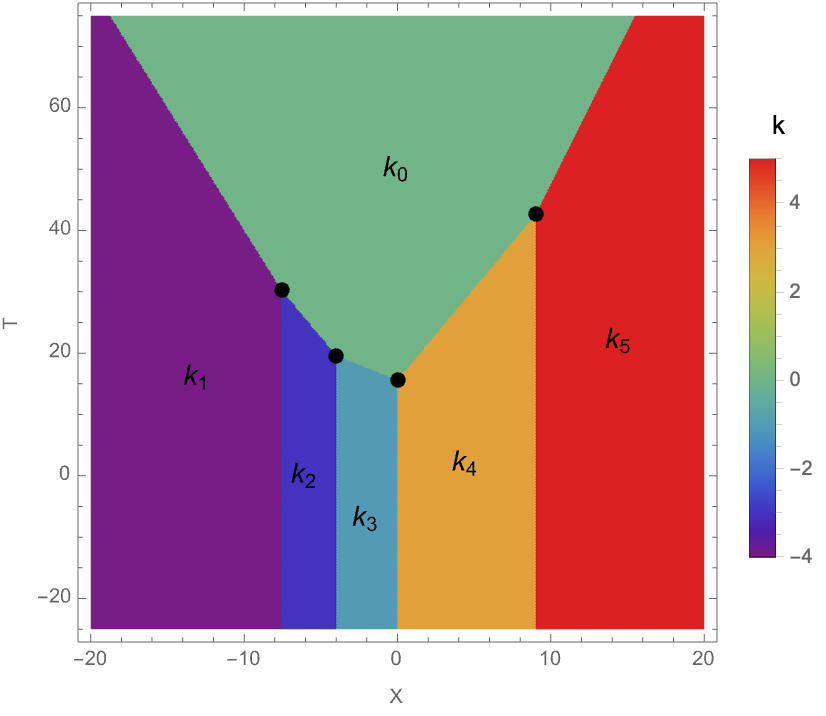}
\caption{\small{Multi-kink solution~\eqref{eq N kink Delta0} with $N=5$. Parameters are set to $k_0=0,  \, k_1=-4, \, k_2=-3,\, k_3=-1,\, k_4=3 \,, k_5=5$,  
	$\delta_0=\delta_1=\ln(3/2), \, \delta_2=8,\, \delta_3=\delta_4=16,\, \delta_5=-2,\, \omega^{(0)}=1$ and $\eta=1$.  \emph{(Left)} solution for $u(X,T)$.   \emph{(Right)} geometric construction identifying the asymptotic ($\eta \to 0$), discontinuous solution.
	}} 
	\label{fig: N5 Delta0}
\end{figure}

Plots can be given to illustrate the implied dynamics at different times, see for instance Figures~\ref{fig: N3 Delta0} and~\ref{fig: N5 Delta0}.   Figure~\ref{fig: N5 Delta0} shows that  the solution is initially, for $T<T_{034}\simeq 15.6$, \emph{approximately stationary}\footnote{Notice that, solution \eqref{eq N kink Delta0} can be written in the form \eqref{sol Delta 0} with $A(X)=\sum_{j=1}^N e^{(k_j X+\delta_j) /\eta}$ and $B(T)=e^{(\omega_0 t + \delta_0)/\eta}$. As a result, the time dependence is visible only when the term $\omega_0 t +\delta_0$ is sufficiently large compared to terms $k_j x +\delta_j$. }, displaying four shocks at rest. 
When $T=T_{034}\simeq 15.6$ a first fission occurs at $X=X_{034}=0$, generating an inner vacuum state and two shocks travelling in opposite directions. Multiple fusions occur until only two shocks survive at large times, i.e. $T \gtrsim 43$. One can verify that, similarly to the case $\Delta\neq 0$,  introducing the mass and momentum densities, $\partial_X u(X,T)$ and $-\partial_T u(X,T)$, respectively,  total mass and momentum are conserved. 
	Explicitly, we have 
	\begin{align}
		P&=- \int_{-\infty}^\infty \, \p_Tu \, dX =   \int_{-\infty}^\infty \,\eta \left[\p_X \left( \, \p_T \ln u \right)\right]\, dX = \eta \, \p_T \ln u  \bigg\rvert_{-\infty}^{+\infty}  = \omega_+ - \omega_{-}
	\\
			M&= \, \, \,\int_{-\infty}^\infty \, \p_X u \, dX =  u \big\rvert_{-\infty}^{+\infty} =  k_+ -k_-\,,
		\end{align}
	where $\pm$ labels the outer states at $X \to \pm \infty$ as before.
	 This holds  also locally at each shock fission/fusion process. 
	Analogously to Eqs.~\eqref{eq:conservation of mass} and \eqref{eq:conservation of momentum}, conservation of mass and momentum in an elementary fusion of two shocks takes the simple form 
	\begin{align}
	(k_i-k_l)+(k_l-k_j)&=k_i -k_j \\
	(\omega_i-\omega_l)+(\omega_l-\omega_j)&= \omega_i -\omega_j\,.
	\end{align} 
 For instance, Figure~\ref{fig: N3 Delta0} displays the multiple fusions of a shock of amplitude (or mass) $m_{02}=k_2-k_0=1$ travelling with velocity $v_{02}=\omega^{(0)}/k_2=1$ with two stationary shocks  of amplitude $m_{21}=k_1 -k_2=2.5$ and $m_{13}=k_3 -k_1=1.5$. In the first collision, a shock of amplitude  $m_{01}=m_{02}+m_{21}=3.5$ is produced, which moves with velocity 
$v_{01}=\frac{m_{02}}{m_{02}+m_{21}} v_{02}=2/7$, hence fulfilling  conservation of mass and momentum.
	
 \section{B\"acklund Transformations for Equation \eqref{eqscaled}}
\label{backlund}

B\"acklund transfomations (BTs) remodel a given nonlinear PDE into some other PDE, thereby relating  the solutions of the two \cite{wolfangbook}. This strategy has proved highly beneficial  to generate answers to nonlinear PDEs, playing a fundamental role in the  growth of activities and interest in soliton theory.  BTs generally link two functions in a system  of first order PDEs, the two functions being a BT if they both satisfy the PDE independently.  
In this Section we deal with problem of finding B\"acklund transformations for the Equation \eqref{eqscaled}.

\subsection{One--parameter B\"acklund transformations}
In this Subsection, we set up BTs for the Equation  \eqref{eqscaled}.  
Aimed at this, we employ a strategy that is similar to that developed for determining BTs in the case of the Bateman-Burgers equation. 
 That is,  let us turn back to the linear equation \eqref{KGeq} pertinent to Equation  \eqref{eqscaled} and consider a special solution $\overline{\phi}$. Then, the corresponding solution for $u$ is given by \eqref{CH X}, i.e. 
$ \overline{u} = \p_X\overline{\phi} / \overline{\phi}$. Now, also the function $\overline{\overline{\phi} } =\p_X \overline{\phi}$ satisfies the equation  \eqref{KGeq}. Therefore, a new  solution will be given  by 
\beq  \overline{\overline{u}} =  \frac{\p_X\overline{\overline{\phi}}}{\overline{\overline{\phi}}} =  \frac{\p_{XX}\overline{\phi}}{\p_X \overline{\phi}} . \label{newsol0} \eeq
Since $ \p_{XX} {\overline{\phi}} =\overline{\phi} \, \p_X \overline{u}+  \overline{u}\, \p_X {\overline{\phi}}  =  \left(\p_X  \overline{u} +  \overline{u}^2\, \right) {\overline{\phi}}$, by combining the previous relations,  one can eliminate the auxiliary function $ {\overline{\phi}}$ to obtain the formula
\beq 
\overline{\overline{u}}  =   \frac{\p_X \overline{u}}{\overline{u}}\, + \, \overline{u} \,\, ,  
\label{eq:BT lambda}
\eeq
as in the Bateman-Burgers case. This is a very special case of B\"acklund transformation, allowing  us to generate a new solution for Eq.  \eqref{eqscaled}  from a known one.
 Formula \eqref{eq:BT lambda} may have, however, a limited use as far as one is interested in smooth functions. For example, it leads immediately to  a singular class of solutions for the case of travelling waves \eqref{eq:TW scaled} in Figure \ref{fig: travelling waves}.
A possible `way out' stands in striving to put forward a  generalisation of the above  line of reasoning  by  imposing the dependency on an auxiliary parameter. The objective is to  bring in a control parameter able to rule out singular solutions if not germane to the phenomenology aforethought for the very specific problem dealt with \eqref{GLMeq}. In particular, we can proceed with this idea by introducing  a parameter $\lambda\in \mathbb{C}$ according to
\beq   {u}^\lambda  =  \frac{\p_X u}{\lambda+ {u}}\, +\, {u}\, \, ,
\label{BT-GLMeq}\eeq
as it can be proved by direct substitution of $u^\lambda$ into Eq.~ \eqref{eqscaled}.

\setcounter{theorem}{3}

 \begin{remark} 
 {\em 
 It is  noteworthy that when we  formulate the problem \eqref{eqscaled} in the form  \eqref{systurho}, the corresponding B\"acklund  transformation reads
\beq  {u}^\lambda  =  \frac{\p_X {u}}{\lambda+ {u}}\, +\, {u}\, , \qquad  {\rho}^\lambda  =  \frac{ \p_X {\rho}}{\lambda+ {\rho}}\, +\, {\rho}\, \,. 
\label{eq:Backlund system}
\eeq
This result  can be proved by direct substitution of the  expressions \eqref{eq:Backlund system} into the system \eqref{systurho}. Then, by using the same system to simplify the coefficients of $\lambda$, one can  verify that they are correctly posed.
}
 \end{remark}

 To make a plain instance of the outcomes from \eqref{BT-GLMeq}, and of the functioning of the introduced auxiliary parameter, 
 we can apply this transformation to the travelling wave solution  \eqref{eq:TW scaled}, so to get 
\beq
u^\lambda-u=\frac{\p_X u(X,T)}{u(X,T)+\lambda}=\frac{\left(K_1^2+\frac{1}{c}\right) \text{sech}^2\left[\sqrt{K_1^2+\frac{1}{c}} (X-c\, T+K_2)\right]}{K_1+ \sqrt{K_1^2+\frac{1}{c}} 
		\tanh \left[\sqrt{K_1^2+\frac{1}{c}} (X-c\, T+K_2)\right]+\lambda } \,\, .
\label{ulambda u}
\eeq
The BT  puts into effect the introduction of another travelling component adding to the initial profile of $u$ and moving with the same velocity, determined by the r.h.s. of \eqref{ulambda u}. Singularities occur if $\lambda\in \left[K_1 -\sqrt{K_1^2+c^{-1}} , K_1+ \sqrt{K_1^2+c^{-1}} \right] $.

\begin{figure}[h!]
	\centering 
	\includegraphics[scale=0.23,angle=0]{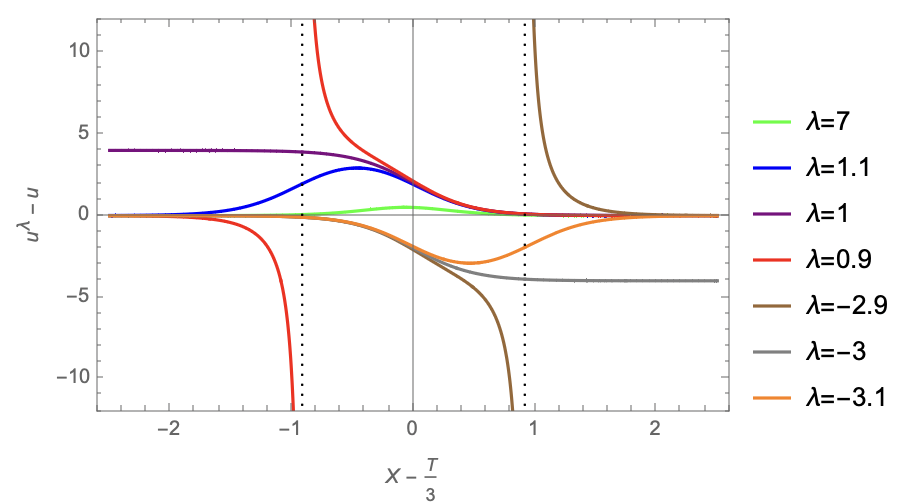}  
	\includegraphics[scale=0.23,angle=0]{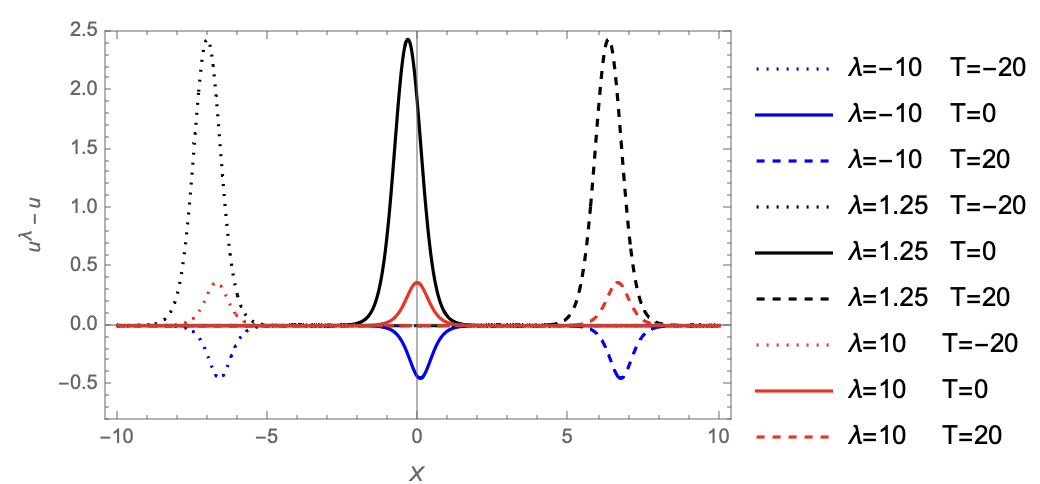} 
	\caption{
		\small{
			Plots of \eqref{ulambda u} for different values of $\lambda$. The seed solution $u$ is the same considered in the first panel of 
			Figure \ref{fig: travelling waves} with $K_1= 1$, $K_2= 0$ and $c = 1/3$. 
{\em (Left)} $u^\lambda-u$ evaluated as function of its argument $X-T/3$ for different values of $\lambda$. 
{\em (Right)} Translation over the $X$-axis of the shape of  $u^\lambda-u$ evaluated as function of $\lambda$ and at distinguished $T$s. 
		}
	}
	\label{TW lambda}
\end{figure}

Figure \ref{TW lambda} shows the variety  of   results for Eq.~\eqref{ulambda u} while varying the parameter $\lambda$.  
The solution $u$ to which the BT \eqref{BT-GLMeq} is applied is the same considered in the first panel of Figure \ref{fig: travelling waves}, namely Eq.~\eqref{eq:TW scaled}  with $K_1= 1$, $K_2= 0$ and $c = 1/3$. The Left panel in Figure \eqref{TW lambda} shows the generation and removal of singularities occurring for such new shape component while tuning the auxiliary parameter $\lambda$,  along with the deformations of logistic sigmoids  generating equal asymptotic limits and flattening the generated extremum. 
For $\lambda=1$ and $\lambda=-3$, for instance, the solution is regular in the whole domain. When $\lambda  \in(-3,1)$ singularities originate (brown and red curves).

For positive (resp. negative) $\lambda$ close to the value $\lambda=1$ (resp. $\lambda=-3$) bumps develop as displayed in the blue curve (resp. orange), but  as  long as the value of $\lambda$ is taken sufficiently far away from the range $[-3,1]$, curves flatten about the horizontal axis, as displayed by the green curve.  The right panel provides examples of the motion for the additional travelling component that adds to $u$ via BT to define the solution $u^\lambda$.

\begin{figure}[h!]
	\centering 
	\includegraphics[scale=0.28,angle=0]{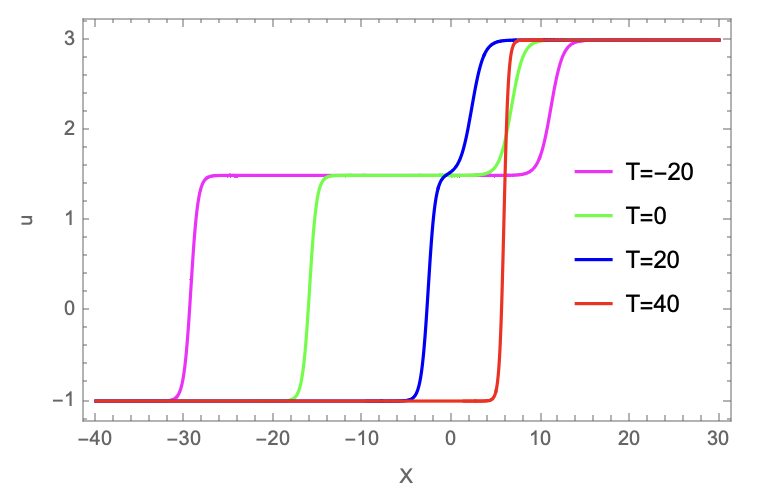}  
	\includegraphics[scale=0.28,angle=0]{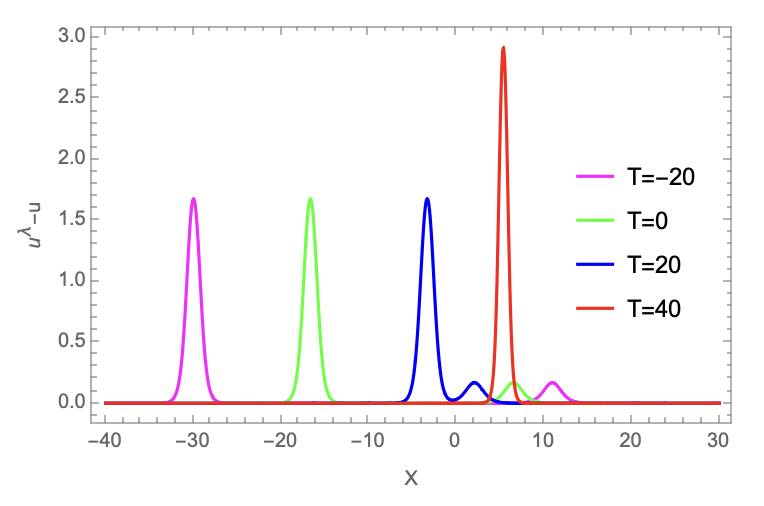}  
	\caption{
		\small{
			Plots of $u$ and $u^{\lambda}-u$ resulting from  \eqref{eq:weight decomposition} and  \eqref{BT-GLMeq}. (\emph{Left}) Evolution of the $N=3$ solution $u$  with parameters $k_1=-1,k_2=1.5,k_3=3, \delta_1=-60,\delta_2=-20,\delta_3=-30$. (\emph{Right}) Evolution of components added to  the seed solution via a BT with parameter $\lambda= 1.1$. 
		}
	}	
	\label{TW N3}	 
\end{figure}

\begin{figure}[h!]
	\centering 
	\includegraphics[scale=0.57,angle=0]{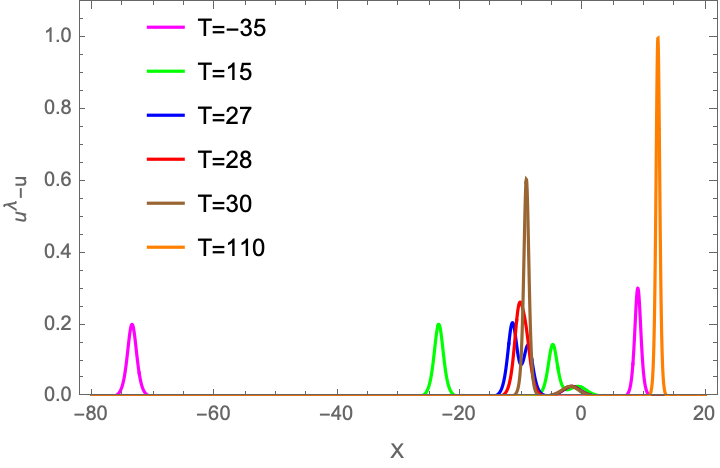}  \quad
	\includegraphics[scale=0.57,angle=0]{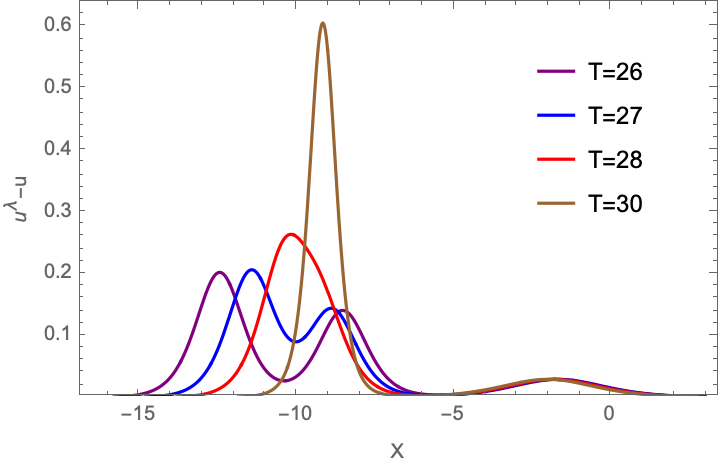} 
	\caption{\small{
			Plots of \eqref{ulambda u} for a multi-kink seed solution $u$ \eqref{eq:weight decomposition} with $N=5$ and parameters $k_j$, $\delta_j$ as in Figure \ref{fig: 5coll}. 
			$\lambda=5$ ensures regularity in the local variable range considered. {\em (Left)} Equation \eqref{ulambda u} at different $T$. {\em (Right)} Close-up of the interaction for $T$ in the range 
			$(26,30)$. The value $T=26$ (purple) was not considered in the left plot for the sake of a better figure clarity.   
		}
	}
	\label{TW N5}
\end{figure}

We can also  provide other examples of solutions obtained by application of  \eqref{BT-GLMeq} by resorting to the kink-type solution
\eqref{eq:weight decomposition} that we derived earlier in Section \ref{multi kink 2}.  Notice that singularities for \eqref{BT-GLMeq}
are prevented provided  that $\lambda\neq -\sum_{j=1}^{N} k_j $.  The plot on the left of Figure \ref{TW N3}  shows  the evolution of a solution  \eqref{eq:weight decomposition} with $N=3$. The solution exhibits, at successive values of $T$,  fusions that   absorb  the intermediate  plateau while progressing towards the direction of increasing $X$ values. The second plot discloses the differences at the same  values of  $T$  between $ u^{\lambda}$ and $u$. Two initial travelling components approach each other, their location being centred about the high-gradient regions and their magnitude  being larger when a higher gradient is required to connect two asymptotic states. The two components ultimately merge, originating a peaked profile moving according to the latest fusion process.  Higher values of $N$ introduce more articulated shapes and interactions, of course. Nevertheless, essential features are preserved for the overall dynamics. This is supported by Figure \eqref{TW N5}, showing the dissimilarities between $ u^{\lambda}$ and $u$ are when in the case $N=5$. A pronounced peak forms for the function 
$ u^{\lambda}-u$ at large $T$ (orange curve in the left plot),  after a cascade  of interactions between the different travelling components.  The interaction occurring for times $26 \leq T \leq 30$ is better evidenced by the close-up in the right panel.  

\subsection{  Two--parameter B\"acklund transformations}
Having obtained BTs in the form  \eqref{BT-GLMeq} and  understood the effects of their auxiliary parameter $\lambda$, we can now go further and analyse the action of BTs involving two parameters by suitable composition of single parameter BTs. Indeed, owing to the Cole-Hopf transform \eqref{CH X} and its derivative w.r.t. $X$, one obtains a generalisation of \eqref{newsol0}, namely
\beq
{u}^\lambda  = \frac{\p_{XX} \phi + \lambda \p_X \phi}{\p_X \phi + \lambda \,\phi}\, \,,
\eeq  
and the linear combination $\p_X \phi + \lambda \, \phi$ is a new solution of the linear problem \eqref{KGeq}. This suggests to take two distinct values of the transformation parameter, say $\lambda_1$ and $\lambda_2$, and use the above relation \eqref{BT-GLMeq} to provide the solutions $u^{\lf \lambda_1, \, \lambda_2\rg}$ and  $u^{\lf \lambda_2, \, \lambda_1\rg}$, which actually coincide. Moreover,  eliminating the explicit presence of the parameters, one can find a superposition formula like 
\beq 
u^{ \lambda_1, \, \lambda_2} = u^{  \lambda_2} + \frac{\lf u^{ \lambda_1} -u\rg u_X^{ \lambda_2}}{\lf u^{ \lambda_1} -u\rg \lf u^{ \lambda_2} -u\rg + u_X} \,\,. 
\label{pseudo-superpos}
\eeq
In terms of the function $\phi$  one obtains the formula
\beq  u^{ \lambda_1, \, \lambda_2} = \frac{ \p_{XXX} \phi
	+ \left(\lambda _1+\lambda
	_2\right) \p_{XX} \phi+\lambda _1 \lambda _2
	\p_X    \phi }{\p_{XX} \phi
	+\left(\lambda _1+\lambda
	_2\right) \p_X \phi  + \lambda _1 \lambda _2
	\phi } \,\,. 
\eeq 
The BTs established above can be  exploited repeatedly and combined to originate further solutions. 
For instance, a Bianchi permutability scheme can be superimposed as follows.  One can start by combining two BTs    generated by  the same seed solution in correspondence of two different parameters. The application of  a second BT with the  interchanged  parameters, and further, the  request of compatibility of the results, leads to a formula which provides a fourth solution in the form of an algebraic  (nonlinear) expression of the previous three ones. 

\begin{figure}[h!]
	\centering 
	\includegraphics[scale=0.216,angle=0]{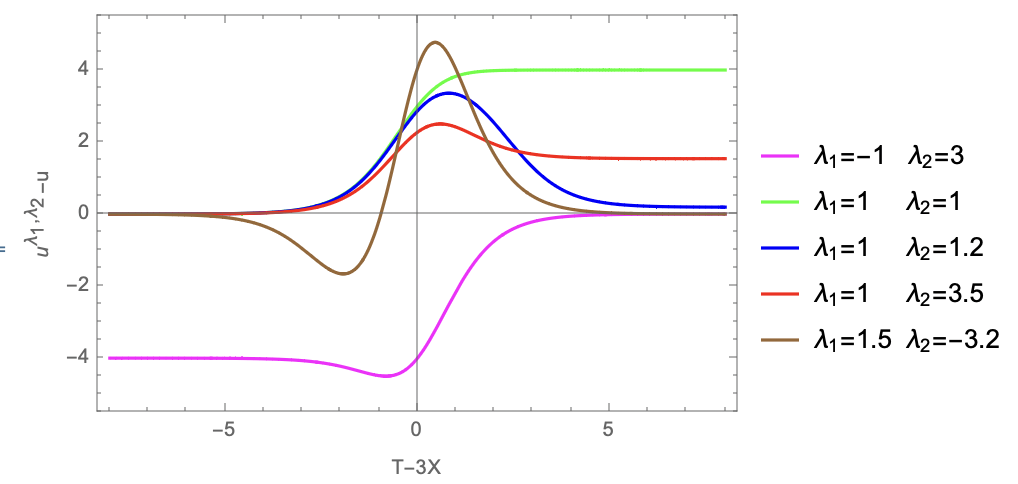}  
	\includegraphics[scale=0.216,angle=0]{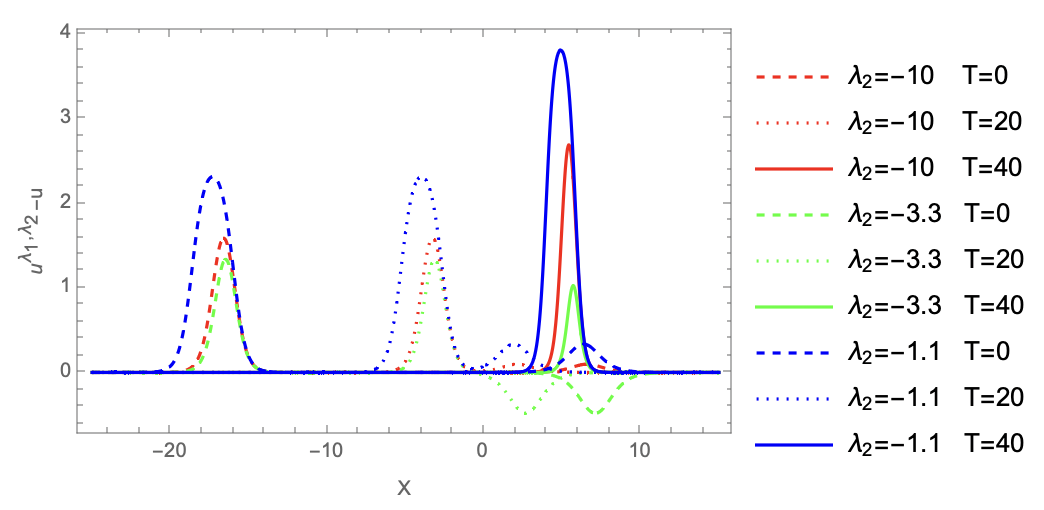}  
	\caption{
		\small{
			Plots of $u^{\lambda_1,\lambda_2}-u$ resulting from \eqref{pseudo-superpos} for different values of  $\lambda_1$ and $\lambda_2$. (\emph{Left}) the seed solution $u$ is the same considered in the first panel of Figures \ref{fig: travelling waves} and \ref{TW lambda}, with $K_1= 1$, $K_2= 0$ and $c = 1/3$. (\emph{Right}) $N=3$ kink solution $u$ depicted in Figure \ref{TW N3}, showing the new features introduced by a two parameter BT with $\lambda_1=1.1$ and  $\lambda_2=-10,-3.5,1.1$. 
		}
	}		 
	\label{fig: BT2}
\end{figure}

We now compare results from the implementation of one- and two-parameter BTs. Let the seed function $u$ be once again the travelling wave
\eqref{eq:TW scaled}.  The application of formula  \eqref{pseudo-superpos} yields
\beq
u^{\lambda_1,\lambda_2}-u=\frac{(1+c \,K_1^2) \{c \,R(Y) \,[R(Y)+(2K_1+\lambda_1+\lambda_2)] +[1-c\lambda_2( 2K_1+\lambda_2)]\} } {
	\cosh^2(Y)
	\{ c\, R(Y) [c(2K_1+\lambda_1+\lambda_2) R(Y)] +[1-c\lambda_2( 2 K_1+\lambda_2)] \}
}
\eeq
where 
\beq
Y=K_1\, \sqrt{K_1^2+\frac{1}{c}} \,\,\ (X-c \,T+K_2) \,\, , \qquad R(Y) =K_1+\lambda_2 +\sqrt{K_1^2 +\frac{1}{c}} \tanh(Y)\,\,.
\eeq
Figure \ref{fig: BT2} shows examples of effects from the application of a two-parameter BT  \eqref{pseudo-superpos} for different values of     $\lambda_1$ and $\lambda_2$. Potential roots of the denominator are displayed at 
\beq
X^*_{1}(T)=\frac{T}{3}-\frac{1}{2} \tanh ^{-1}\left(\frac{\lambda_2+1}{2}\right) \,\, , \qquad 
X^*_{2}(T)=\frac{T}{3}-\frac{1}{2}\tanh ^{-1}\left[\frac{\lambda_1 \lambda_2+\lambda_1+\lambda_2+5}{2 (\lambda_1+\lambda_2+2)}\right] \,\, .
\eeq
However, in Figure  \ref{fig: BT2} we  have avoided to take into account singular cases. 
Effects on the travelling wave solution \eqref{fig: travelling waves} have been delved into first. The  $u^{\lambda_1,\lambda_2}-u$ consists  of fixed profiles travelling. Depending on the values of $\lambda_1$ and $\lambda_2$,  extrema can be generated and asymptotes can be affected (Left panel). The application  of formula \eqref{pseudo-superpos} in respect to the solution for $N=3$  in Eq. \eqref{eq N kink Delta0} is shown in the Right panel. Our findings  validate  arguments parallel to those inferred while referring to the action of a BT \eqref{ulambda u} on the same multi-kink solution.

\section{Conclusions and future perspectives}

 The present communication has been devoted to the study of the nonlinear non-evolutive PDE \eqref{GLMeq}, which can be seen as a generalised Fokker-Planck equation encoding a density-dependent diffusivity, and nontrivial nonlinearity and  viscous effects. The equation, which was introduced in \cite{GLM} for the description of real gases within a mean-field framework,  is potentially applicable to general circumstances where  a conservation law is foreseen with a current that depends (linearly) also on first derivative of both spatial- and time-type local variables, in fact.

A first measure in  our  analysis  has been to identify two sets of coordinates through the implementation of  two distinct linear transformations on the dependent and independent variables, i.e. Eqs. \eqref{CRvisco} and \eqref{CRefScaling}. Such transformations allows us to make the analysis of the resulting PDEs independent on the structural parameters of the model equation under study, provided that  a discerning constraint is assessed ($\Delta\neq 0$ or $\Delta=0$). This clearly led to a simplification of the differential problem and a more favourable starting point for the analysis of its integrability features.  
 
We worked out a concrete study with reference to travelling wave solutions and presented a comprehensive clarification of dynamics for a class of naturally discerned multi-kink solutions, that incidentally turned out to be characterised by a simple reciprocal dispersion relation. 
By first restoring the role of the small parameter $\eta$ originally associated with  viscosity in the conservation law \eqref{GLMeq}, we provide an interpretation of multi-kink solutions, in the limit $\eta \to 0$, in terms of classical viscous shock waves.  Our analysis and description of multi-kink solutions relies on a scheme that entails the  partitioning of the space of independent variables into disjoint regions $\mathcal{R}_j$ where each of the exponential terms $e^{\theta_j /\eta}$ contributing to the sum   is  dominant (see Eq. \eqref{eq:n term phi}).  
The small viscosity  regime  of the multi-kink solution results into approximating the sum with the dominant term in each region and is analogous to the so-called tropical limit (see Eq. \eqref{eq:geometric construction}) \cite{kato,Kodama Williams,Dimakis}. 
The lines which separate the disjoint regions $\mathcal{R}_j$ constitute the  fundamental information about shock trajectories and their interaction,  their intersections providing the collision location and time (see explicit formula \eqref{vertex}).
We identified conserved quantities, such as mass and momentum,  for the differential equations under study in the two different cases and analyse\FG{d} the interaction between shocks as scattering processes among particles. This allowed us to establish  a shock--particle duality, which was detailed with the aid of representative examples. In particular,  we showed that conservation of mass and momentum for the underlying equations can be formulated as analogous conservation  laws for particles (see  Eqs. \eqref{eq:conserv of mass} and \eqref{eq:conserv of momentum}). We also proved that, sufficiently far  from the interaction, the multi-kink solution is approximated as a collection of travelling waves in the form of single kinks, like those determined in Section \ref{section travelling}, hence demonstrating that there is no phase-shift resulting from the kink-kink interaction.

Our understanding of  integrability features of Eq. \eqref{GLMeq} has been later enhanced by the devising of B\"acklund transformations (BTs) depending  on auxiliary parameters, whose effect has been argued  showing  the  generation of   travelling components adding to the shape of the \emph{naked} solution.   
 To achieve this, we have discussed examples of the applications of one- and two- parameter BTs. 

Our analysis has introduced new useful insights into solutions and properties of Equation  \eqref{GLMeq}.
Nonetheless, several lines of investigation can be yet pursued related to its solvability and integrability.
These comprise the existence and  analysis of solutions subject to other relevant (dynamical or boundary) conditions.  Among others,   
special solutions which can be written in terms of theta functions may be sought, following the procedures employed for the Bateman-Burgers \cite{parker} or other PDEs (see
e.g.  \cite{bertola, Kodama-theta} and Refs. therein). 
Other fundamental directions can also be explored, such as those arising from our findings in the context of the Hirota method, 
BTs and Lax pairs. For instance, the implementation of the Hirota method accounting for the improvements related to the introduction of combinatorial methods \cite{Gilson,Ma}
may be assessed.    Moreover,  a systematic study of higher order (generalized)  symmetries \cite{vinogradov,olver} is also missing. 
By all such methods one may be led to retrieve novel classes of  solutions (maybe made of distinct interacting components as in \cite{Zarmi}),
design different B\"acklund transformations,  seek for insights concerning a  Lax pair formulation and identify infinitely many commuting conservation laws for the nonlinear PDE \eqref{GLMeq}. 
Yet another direction of study can be the discretisation of the model and the subsequent queries raised and application of devoted tools,   e.g. the  Inverse Spectral Transform  following the guidance developed in  \cite{LRB}. 
 Last but not least, extensions to higher dimensions  of \eqref{GLMeq}  could be constructed. Beyond presumably offering  a richer phenomenology and new classes of solutions, as usually it happens in nonlinear soliton equations in higher dimensions (see e.g. \cite{konopelchenko2 } and Refs. therein),
these might either establish or strengthen the connection with known and widely studied higher dimensional models, e.g. reductions from Navier-Stokes equations other than  the Bateman-Burgers one, higher order Fisher or Fokker-Planck equations and so forth \cite{benjamin, wang tang, Bao, Du}.

Finally, we wish to highlight insights and  perspectives offered by our study in the originating context of Eq. \eqref{GLMeq}. Indeed, the integrability structure of the model discussed in this work shares similarities with  a plethora of one- and multi-component  models  recently studied in the realm of statistical physics (see for instance \cite{Moro annals,Biondini,De Matteis biaxial} to cite a few). One of the key--features of all those models is the fact that order parameters, as functions of thermodynamic variables, fulfil  nonlinear C-integrable conservation laws of viscous type. The C-integrability property implies that the  partition function  of the model obeys to a linear differential identity, playing incidentally the role of a linearising potential function linked to the order parameter through a Cole-Hopf transformation. As  a result,  the partition function is  a sum of exponential terms of the form $e^{\theta_j/\eta}$, with the parameter $\eta=1/N$, $N$ being the  number of particles in the statistical ensemble. In that context, each function $\theta_j$ is full determined by  the energy and statistical weight of each allowed macrostate of the system. The results presented in this  work imply that the thermodynamic limit $N \to \infty$ for models with a  countable  number of macrostates could be seen as a tropical limit, hence establishing links with the work \cite{Angelelli}, where the Authors shed light on the relevance of the tropical limit in statistical physics, intended as the limit $k\to 0$, with $k$ being the Boltzmann constant. 
We feel that our approach and results may lead to advancements in the asymptotics of order parameters for a large class of C-integrable models 
and potentially describe  statistical systems exhibiting  complex cascades of phase transitions.

\subsubsection*{Acknowledgments}
FG and GL would like to thank the Isaac Newton Institute for Mathematical Sciences, Cambridge, for support and hospitality during the programmes ``Dispersive Hydrodynamics: mathematics, simulation and experiments, with applications in nonlinear waves'' and  ``Emergent phenomena in nonlinear dispersive waves'' where work on this paper was undertaken. This work was supported by EPSRC grant no EP/R014604/1. FG also acknowledges the hospitality of the Lecce’s division of INFN and of the Department of Mathematics and Physics ’Ennio De Giorgi’ of the University of Salento. GL and LM acknowledge the hospitality of the School of Mathematics and Statistics of the University of Glasgow. INFN IS-MMNLP partially supports GL and LM. GL is also partially supported by  INFN IS-GAST. 
Authors are indebted to C Gilson, Y Kodama, B Konopelchenko and A Moro for useful discussions and remarks. 

\subsubsection*{Competing interests} 
 The author(s) has/have no competing interests to declare. 
 \subsubsection*{Licence} 
 For the purpose of open access, the authors have applied a Creative Commons Attribution (CC BY) licence to any Author Accepted Manuscript version arising from this submission.

 \end{document}